\documentclass[iop]{emulateapj}

\shorttitle{Calibration of the Mixing-Length Theory for Convective White Dwarf Envelopes}
\shortauthors{Tremblay et al.}

\begin{document}

\title{CALIBRATION OF THE MIXING-LENGTH THEORY FOR CONVECTIVE WHITE
  DWARF ENVELOPES}

\author{P.-E.~Tremblay\altaffilmark{1}}
\affil{Space Telescope Science Institute, 3700 San Martin Drive,
  Baltimore, MD, 21218, USA}
\email{tremblay@stsci.edu}

\author{H.-G.~Ludwig}
\affil{Zentrum f\"ur Astronomie der
  Universit\"at Heidelberg, Landessternwarte, K\"onigstuhl 12, 69117
  Heidelberg, Germany}
  
\author{B.~Freytag}
\affil{Department of Physics and
  Astronomy at Uppsala University, Regementsv\"agen 1, Box 516,
  SE-75120 Uppsala, Sweden}
\affil{Centre de Recherche
  Astrophysique de Lyon, UMR 5574: CNRS, Universit\'e de Lyon, \'Ecole
  Normale Sup\'erieure de Lyon, 46 all\'ee d'Italie, F-69364 Lyon
  Cedex 07, France}
  
\author{G.~Fontaine}
\affil{D\'epartement de Physique,
  Universit\'e de Montr\'eal, C. P. 6128, Succursale Centre-Ville,
  Montr\'eal, QC H3C 3J7, Canada}
  
\author{M.~Steffen}
\affil{Leibniz-Institut
  f\"ur Astrophysik Potsdam, An der Sternwarte 16, D-14482 Potsdam,
  Germany}
  
\and

\author{P.~Brassard}
\affil{D\'epartement de Physique,
  Universit\'e de Montr\'eal, C. P. 6128, Succursale Centre-Ville,
  Montr\'eal, QC H3C 3J7, Canada}

\altaffiltext{1}{Hubble Fellow}

\begin{abstract}
A calibration of the mixing-length parameter in the local
mixing-length theory (MLT) is presented for the lower part of the
convection zone in pure-hydrogen atmosphere white dwarfs. The
parameterization is performed from a comparison of 3D CO5BOLD
simulations with a grid of 1D envelopes with a varying mixing-length
parameter. In many instances, the 3D simulations are restricted to the
upper part of the convection zone. The hydrodynamical calculations
suggest, in those cases, that the entropy of the upflows does not
change significantly from the bottom of the convection zone to regions
immediately below the photosphere. We rely on this asymptotic entropy
value, characteristic of the deep and adiabatically stratified layers,
to calibrate 1D envelopes. The calibration encompasses the convective
hydrogen-line (DA) white dwarfs in the effective temperature range
$6000 \leq T_{\rm eff}$ (K) $\leq 15,000$ and the surface gravity
range $7.0 \leq \log g \leq 9.0$. It is established that the local MLT
is unable to reproduce simultaneously the thermodynamical, flux, and
dynamical properties of the 3D simulations. We therefore propose three
different parameterizations for these quantities. The resulting
calibration can be applied to structure and envelope calculations, in
particular for pulsation, chemical diffusion, and convective mixing
studies. On the other hand, convection has no effect on the white
dwarf cooling rates until there is a convective coupling with the
degenerate core below $T_{\rm eff} \sim$ 5000~K. In this regime, the
1D structures are insensitive to the MLT parameterization and converge
to the mean 3D results, hence remain fully appropriate for age
determinations.
\end{abstract}


\section{INTRODUCTION}

In late-type stars, giants, and cool white dwarfs, the convective
outer envelope has a significant impact on the observed properties.
The physical principles explaining convective energy transport in
stars are well understood, although the non-local and turbulent nature
of convection has delayed the development of precise models for
convective stellar layers. The mixing-length theory \citep[][hereafter
  MLT]{MLT} has proven rather successful despite presenting a very
simple description of convection. In this picture, the condition that
distinguishes between convective and stable layers is the
Schwarzschild criterion, and the convective efficiency, the ratio of
convective and radiative fluxes, is computed from local quantities. In
the super-adiabatic convective layers that define the atmosphere of
most stars, the predicted convective efficiency is very sensitive to
the underlying model describing the radiative energy losses, the
lifetime, and the geometrical shape of individual convective
structures. These quantities are not well constrained by the MLT and
must be calibrated from observations.

In recent years, three dimensional (3D) radiation hydrodynamical (RHD)
simulations have provided predictions for the surface convection that
are in very good agreement with the observed solar granulation
\citep[see, e.g.,][]{wedemeyer09}. Furthermore, various studies relied
on 3D RHD simulations to improve the predicted photospheric structures
and spectroscopic abundance determinations for the Sun and other stars
\citep{asplund09,caffau11,scott14a,scott14b}. In addition to a better
representation of the surface inhomogeneities, 3D model atmospheres
feature non-local effects, such as the so-called top overshoot layers,
which are completely missing in local 1D MLT models
\citep{unno57,ludwig02,nordlund09,freytag10,tremblay13c}.

The deep convection zone, where the stratification becomes essentially
adiabatic, is not sensitive to the convection model. It is however the
entropy jump in the super-adiabatic layers that completely defines the
asymptotic entropy value of the deep, adiabatically stratified
structure, hence also the depth of the convection zone. One
possibility to model these layers is to rely on RHD simulations to
determine the asymptotic entropy value for the deep convection zone
\citep{steffen93,ludwig99}. This arises from the prediction that
upflows formed at the base of the convection zone follow an adiabat
almost up to the visible surface \citep{stein89}. The 1D MLT envelopes
are then calibrated from the multi-dimensional asymptotic entropy, a
technique that has been employed for late-type stars and giants
\citep{ludwig99,ludwig08}. The calibrated 1D structures nevertheless
neglect the overshoot layers predicted at the base of non-local
convection zones \citep{bohm63,chan89,skaley91,freytag96}, which for
deep convective envelopes, impacts the convective mixing into the
nuclear burning core. In this work, we are interested in the
calibration of 1D envelopes of DA white dwarfs with a pure-hydrogen
atmosphere. All currently available white dwarf structures rely on the
local MLT with a fixed parameterization \citep[see,
  e.g.,][]{tassoul90,fontaine01,renado10,salaris10}.

Surface granulation in DA white dwarfs is qualitatively very similar
to that seen in the Sun and stars \citep{tremblay13b}, albeit with
shorter lifetimes and smaller characteristic sizes, which are roughly
inversely proportional to gravity. Convective instabilities due to
hydrogen recombination develop in the atmosphere of these
pure-hydrogen stellar remnants at $T_{\rm eff} \sim 18,000$~K,
although convective energy fluxes only become significant at $T_{\rm
  eff} \sim 14,000$~K for $\log g = 8$. The convection zone eventually
grows to sub-photospheric, and essentially adiabatic layers, at
slightly lower effective temperatures.  White dwarfs in the range
$14,000 \gtrsim T_{\rm eff}$ (K) $\gtrsim 8000$ have super-adiabatic
photospheric layers where the 1D MLT parameterization has a strong
influence on the predicted thermal structures and spectra
\citep{bergeron92,koester94,bergeron95}. \citet{tremblay13c} recently
demonstrated that the local 1D MLT is unable to reproduce the mean
photospheric structure of 3D simulations, and that shortcomings in the
1D MLT are responsible for the spurious high $\log g$ values
previously derived from spectroscopic observations of cool convective
white dwarfs \citep{bergeron90}. In particular, the top overshoot
region was found to have a crucial impact on the spectroscopic
predictions.

The convection zone in DA white dwarfs remains limited to the thin
hydrogen envelope until it reaches the degenerate core at $T_{\rm eff}
\sim 5000$~K, or mixes with the underlying helium layer if the total
gravitationally stratified mass of hydrogen is less than about
$10^{-6} M_{\rm H}/M_{\rm tot}$ \citep{tassoul90}. Before one of these
events takes place, the cooling process is regulated by the radiative
interface layer just above the largely isothermal degenerate core,
which is in some sense the bottle-neck for the energy transport. The
evolutionary calculations converge to the so-called radiative zero
solution, hence they are insensitive to the details of the convection
model \citep{fontaine76}, which is unlike earlier evolutionary stages
\citep[see, e.g.,][]{freytag99}. The situation is different below
$T_{\rm eff} \sim 5000$~K, where the cooling rates are directly
impacted by the convecting coupling between the interior thermal
reservoir and the radiating surface. In this temperature range,
however, the super-adiabatic peak has a negligible amplitude, or in
other words, the full convection zone has an essentially adiabatic
structure which does not depend on the MLT parameterization. As a
consequence, the cooling ages predicted from current 1D evolutionary
sequences are not expected to be impacted by 3D effects. However, the
convection zone has an {\it indirect} effect on observed ages, since
they are often derived from spectroscopically determined atmospheric
parameters that are modified by 3D effects.

There are a number of cases where 3D effects on structures are
expected to have a {\it direct} impact. Non-adiabatic pulsation
calculations depend critically on the structure of the convective
layers, especially for the determination of the edges of the ZZ Ceti
instability strip of pulsating DA white dwarfs
\citep{fontaine94,gautschy96,grootel12}. Chemical diffusion
applications \citep{paquette86,pelletier86,dupuis93} and convective
mixing studies \citep[see, e.g.,][]{chen11} also depend critically on
the size and especially the dynamical properties of the convection
zone, e.g. the root-mean-square (RMS) vertical velocity in the
convective overshoot layers at the base of the convection zone
\citep{freytag96}. In order to characterize white dwarfs accreting
disrupted planets, it is likely important to account for the currently
neglected convective overshoot \citep{koester09}. The total mass of
the chemical elements mixed in the convection zone (hereafter mixed
mass), and to a lesser degree their relative abundances, depend on how
rapidly these elements diffuse in the deep overshoot region. Through
the remaining of this work, overshoot refers only to the region at the
base of the convection zone, since the top overshoot layers have no
direct relevance for white dwarf envelope and structure models.

This study proposes a calibration of the MLT free parameters for the
size of the convection zone in 1D envelopes of DA white dwarfs from a
comparison with CO$^{5}$BOLD 3D simulations previously computed for
spectroscopic applications \citep{tremblay13c}. We emphasize that our
proposed calibration has little in common with the spectroscopic
parameterization of the MLT. In both cases, the free parameters of the
MLT are employed to mimic specific properties of the mean 3D
simulations and mean 3D spectra, respectively, rather than to describe
the more general underlying nature of convection. In Section~2, we
introduce our grid of 3D simulations and 1D envelope models. We follow
in Section~3 with definitions for the sizes of non-local convection
zones. In Section~4, we compare 1D and 3D models in order to propose
and discuss a MLT parameterization in Section~5. We conclude in
Section~6.
 
\section{WHITE DWARF MODELS}

\subsection{3D Model Atmosphere Simulations}

We rely on CO$^{5}$BOLD 3D simulations that were presented in earlier
works \citep[][hereafter TL13a, TL13b, and TL13c,
  respectively]{tremblay13a,tremblay13b,tremblay13c}. The 70
simulations cover the range $6000 \leq T_{\rm eff}$ (K) $\leq 15,000$
and $7 \leq \log g \leq 9$ (see Appendix A of TL13c). While TL13c
reviewed the predicted spectral properties drawn from these
simulations, which mostly depend on the uppermost regions of the
convection zone, the study presented here reports on the overall
properties and lower parts of convection zones. The natural starting
point is therefore the comparison of 3D and 1D structures at $\log g =
8$ presented in TL13a. We have demonstrated that sequences at
different surface gravities possess rather similar properties (TL13b,
TL13c), largely because 3D effects depend mostly on the local density,
and the same range of densities is found at all surface gravities,
albeit with a shift in $T_{\rm eff}$.

The numerical setup of the 3D model atmospheres is described in detail
in TL13a, and more broadly in \citet{freytag12} in terms of the
general properties of the code. We provide a brief overview in this
section. The 3D simulations rely on an equation-of-state (EOS) and
opacity tables that are computed with the same microphysics as that of
standard 1D model atmospheres \citep{tremblay11}. We employed a grid
of $150\times150\times150$ points in the $x,y,$ and $z$ directions,
where $z$ is used for the vertical direction and points towards the
exterior of the star. The grid spacing in the $z$ direction is
non-equidistant and the total horizontal extent is chosen in order to
have about 10 granules at the surface. The structure of the deep
convection zone is largely determined by the radiative energy losses
in the photosphere, which also fix the $T_{\rm eff}$ of a
simulation. As suggested by \citet{brassard97}, \citet{hansen99}, and
\citet{fontaine01}, non-gray atmospheres are an essential boundary
condition for precise envelope calculations. The 3D simulations solve
the non-gray radiative transfer using 8 to 13 opacity bins, which has
proven adequate for spectroscopic applications (TL13c).  This setup is
likely more than sufficient for a comparison with 1D structure
calculations which are less sensitive to the optically thin layers.

The implementation of boundary conditions is described in detail in
\citet[][see Sect.~3.2]{freytag12}. In brief, the lateral boundaries
are periodic, and the top boundary is open to material flows and
radiation.  We rely on bottom conditions that are either open or
closed to convective flows. The lower boundary is closed (hereafter
{\it closed simulations}) when the vertical extent of the convection
zone can be fully included in the simulation. This is the situation
for the 3D simulations with $T_{\rm eff} \gtrsim 10,500, 11,500,
12,000, 13,000$, and 14,000 K, for $\log g =$ 7.0, 7.5, 8.0, 8.5, and
9.0, respectively. In those cases, we impose zero vertical velocities
at the bottom, and a radiative flux is injected from below.

\begin{figure}[h!]
\epsscale{0.8}
\plotone{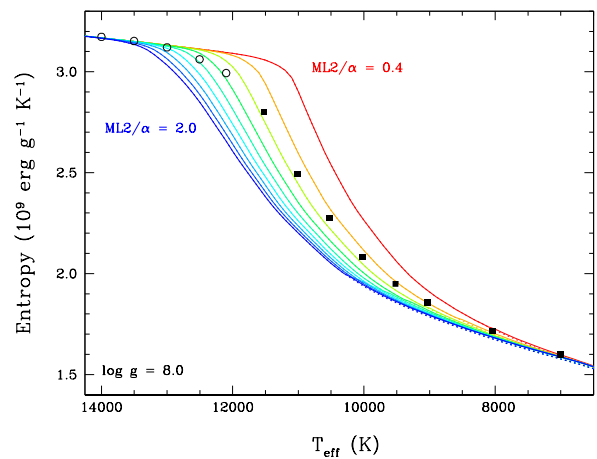}
\caption{Entropy at the bottom of the convection zone as a function of
  $T_{\rm eff}$ for DA white dwarf envelopes at $\log g = 8$. The 3D
  results are shown in black, with the $\langle {\rm 3D}\rangle$
  entropy extracted directly at the Schwarzschild boundary for
  closed simulations (open circles, see Sect.~4.1), and asymptotic
  $s_{\rm env}$ values for open simulations (filled squares, see
  Sect.~4.2). We also display 1D sequences (solid lines, see
  Sect.~2.3) with the MLT parameterization varying from ML2/$\alpha$ =
  0.4 (red) to 2.0 (blue) in steps of 0.2 dex. Additionally, we
  present sequences where gas degeneracy effects are neglected (dotted
  lines), which largely follow the former sequences.
 \label{fg:f_entropy}}
\end{figure}

For cooler simulations, the bottom layer is open to convective flows
and radiation (hereafter {\it open simulations}), and a zero total
mass flux is enforced. We specify the entropy of the ascending
material to obtain approximately the desired $T_{\rm eff}$ value (an
indirect quantity computed from the resulting emergent flux of the
simulation). Figure~\ref{fg:f_entropy} shows that the entropy from 1D
envelopes (see Sect.~2.3) at the lower boundary of the convection zone
increases monotonically with $T_{\rm eff}$. Convection is essentially
adiabatic in deep convection zones, and the entropy value in the lower
part of the convection zone is assumed to be the same as that of the
upflows at the bottom of the simulations (see Sect.~2.2).

In all models, the top boundary reaches a space- and time-averaged
value of no more than a Rosseland optical depth of $\tau_{\rm R} \sim
10^{-5}$. The bottom layer was generally fixed at $\tau_{\rm R} =
10^{3}$, well below the photosphere, i.e. the line-forming regions. A
few models were extended to deeper layers when the bottom of the
convection zone was too close to the simulation boundary. We cover at
least $\sim$3 pressure scale heights ($H_{\rm P}$) below the unstable
regions when the bottom of the simulation is closed to mass flows.

\subsection{Properties of the Deep Convection Zone}
  
\begin{figure}[h!]
\epsscale{0.8}
\plotone{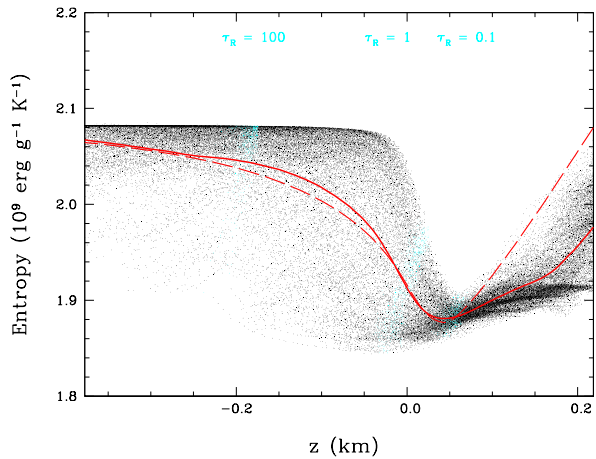}
\caption{Local 3D entropy values (black dots) as a function of
  geometrical depth for a subset of a simulation at $T_{\rm eff} =$
  10025~K and $\log g = $ 8.  The $\langle {\rm 3D}\rangle$ entropy
  profile, averaged over constant geometrical depth, is shown with a
  red solid line.  We also display the 1D entropy (dashed red line)
  with the MLT parameterization calibrated from the 3D simulation
  (ML2/$\alpha$ = 0.69, see Table~2). We highlight $\tau_{\rm R}$
  values at 100, 1.0, and 0.1 (cyan points, values identified in the
  legend) as a guide. The asymptotic 3D entropy value $s_{\rm env}$ is
  2.082$\times$10$^9$ erg g$^{-1}$ K$^{-1}$.
 \label{fg:f_exemple}}
\end{figure}

The physical conditions at the bottom of convection zones can be
extracted from 3D simulations even if we do not simulate the full
zones. We rely on the technique presented in \citet{ludwig99}, for
which a demonstration is shown in Figure~\ref{fg:f_exemple} for a DA
simulation at $T_{\rm eff} = 10,025$~K and $\log g = 8$. We present
the local 3D values of the entropy in convective structures (black
dots) as a function of geometrical depth with the stellar surface on
the right-hand side. We also display the average entropy profile over
constant geometrical depth (solid red line). We observe significant
entropy fluctuations at all depths, although there is a constant {\it
  asymptotic} upper limit, hereafter $s_{\rm env}$. According to the
scenario developed in \citet{stein89} and \citet{ludwig99}, the gas in
central regions of broad ascending flows is still thermally isolated
from its surroundings until it reaches layers immediately below the
photosphere. In other words, convective upflows keep an imprint of the
physical conditions at the bottom of the convection zone. The averaged
3D entropy, on the other hand, is not a conserved quantity due to
radiative losses and the presence of downdrafts created in the
photosphere.

The above technique only applies if the center of upflows remains
adiabatic, hence a minimum requirement is that the conditions at the
bottom of the convection zone are adiabatic. We have observed that the
adiabatic transition takes place when the bottom of the convection
zone reaches layers deeper than $\tau_{\rm R} \sim 10^{3}$. For all of
our simulations with an open bottom, we can recover an asymptotic
value. For closed simulations, there is no significant entropy plateau
since conditions are never adiabatic, although in those cases we can
directly extract the properties at the bottom of the convection zones.

We also overlay in Figure~\ref{fg:f_exemple} the 1D model atmosphere
with the MLT calibrated from a comparison of $s_{\rm env}$ with a grid
of 1D envelopes (see Section 4.2). The 1D model atmospheres and
envelopes calibrated is this way are only meant to recover the
conditions at the bottom of the convection zone, although by
construction they also provide an accurate mean structure for the
essentially adiabatic parts of the convection zone. On the other hand,
there is no guarantee that the calibrated 1D models will provide a
good match to the mean 3D stratification in super-adiabatic
layers. Fortunately, in the case of white dwarfs in contrast to
main-sequence stars, the super-adiabatic layers have little direct
impact on applications that require the use of 1D envelopes. As it was
custom until now, it is generally sufficient to employ 1D envelopes
where the MLT parameterization is based on the deep layers, and rely
on a different set of models, e.g. 3D simulations, for atmospheric
parameter determinations. An inspection of Figure~\ref{fg:f_exemple}
demonstrates that if needed, a connection of the 1D and mean 3D
structures at large depth could also be a fairly good approximation.

\subsection{1D Envelope Models}

For the purpose of this work, we computed 1D envelopes relying on the
MLT for the treatment of convection, similar to those presented in
\citet{fontaine01} and \citet{grootel12}. The models employ the ML2
treatment of MLT convection \citep{bohm71,tassoul90} and an EOS for a
non-ideal pure-hydrogen gas \citep{saumon95}. Realistic non-gray
temperature gradients are extracted from detailed atmospheric
computations and employed as upper boundary conditions
\citep{brassard97,grootel12}. The non-gray effects on the size of the
convection zone are shown in Figure~5 of \citet{grootel12}. In order
to compare the envelopes to 3D simulations, we have varied the mixing
length to pressure scale height ratio\footnote{ML2/$\alpha$ has the
  same functional form as the more commonly used $\alpha_{\rm MLT}$
  for stars but it also specifies the choice of auxiliary
    parameters of the MLT formulation \citep{ludwig99}.} ML2/$\alpha
= l/H_{\rm P}$ from values of 0.4 to 2.0 in steps of 0.2. ML2/$\alpha$
is selected as a proxy for all MLT free parameters since changes in
the other parameters have similar effects on the structures. We use
the same range of surface gravities and $T_{\rm eff}$ (steps of
0.5\,dex and 100\,K, respectively) as our set of 3D calculations.

\begin{figure}
\epsscale{0.8}
\plotone{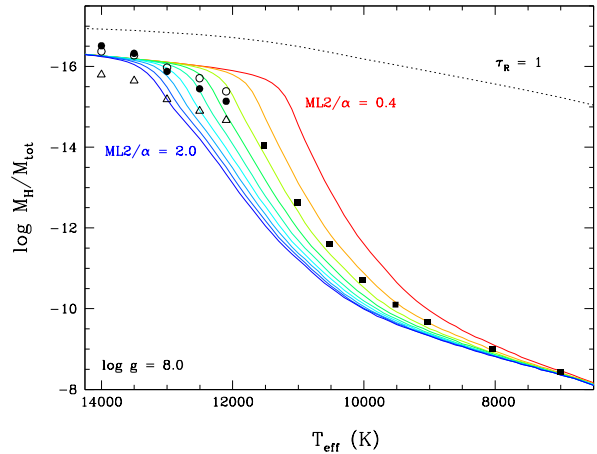}
\caption{Mass of hydrogen integrated from the surface ($M_{\rm H}$)
  with respect to the total stellar mass (logarithmic value) as a
  function of $T_{\rm eff}$ for DA envelopes at $\log g = 8$. The 3D
  results are shown with black symbols using different definitions for
  the bottom of the convection zone (see Sect.~3 and~4). For closed
  simulations, we consider the Schwarzschild boundary (open
  circles), the flux boundary (filled circles), and a $v_{\rm z,rms}$
  decay of 1 dex (open triangles) below the value at the flux
  boundary. For open 3D simulations, the filled squares represent the
  values calibrated by matching $s_{\rm env}$ with the 1D entropy at
  the bottom of the convection zone. We also display 1D sequences
  (solid lines) with the MLT parameterization varying from
  ML2/$\alpha$ = 0.4 (red) to 2.0 (blue) in steps of 0.2 dex. The
  bottom of the stellar photosphere ($\tau_{\rm R} = 1$, 1D
  ML2/$\alpha$ = 0.8), which roughly coincides with the top of the
  convection zone, is represented by a dotted black line.
 \label{fg:f_mass}}
\end{figure}

From the 1D envelopes we have extracted the physical conditions at the
bottom of the convection zone. Figure~\ref{fg:f_mass} depicts the
hydrogen mass integrated from the surface ($M_{\rm H}$), with respect
to the total white dwarf mass, for the $\log g = 8$ case. Clearly, the
MLT parameterization has a strong effect on the size and mass included
in the convection zone at intermediate temperatures, where the
atmospheric layers are super-adiabatic.

To ensure that we share a common entropy zero point in all
calculations, we computed all entropy values using the same EOS as the
3D simulations, based on the \citet{hm88} non-ideal EOS, where we have
also accounted for partial degeneracy. The entropy values at the
bottom of the convection zone are shown in Figure~\ref{fg:f_entropy}
(solid lines), along with additional sequences where we have neglected
partial degeneracy (dotted lines). The degeneracy effects are very
small in the convection zone ($\eta < 0$, where $\eta kT$ is the
chemical potential of the free electrons). This is largely due to the
fact that the degeneracy level is constant for an adiabatic
process. For the {\it essentially} adiabatic structure of cool white
dwarf convection zones, degeneracy is changing very slowly as a
function of depth \citep[see Eq. (13) of][]{bohm68}. Furthermore,
degeneracy effects are still negligible at the lower $T_{\rm eff}$
limit where the calibration of ML2/$\alpha$ is performed in this work
(see Section 5.1).

Our proposed calibration of the MLT is performed by comparing 3D
simulations to 1D envelopes. We also rely on 1D MLT model atmospheres
\citep{tremblay11} for illustrative purposes in cases where we display
a detailed comparison of 1D and mean 3D stratifications as a function
of depth. The 1D model atmospheres and envelopes provide very similar
results, within a few percent, below the photosphere.

\section{DEFINITION OF CONVECTIVE LAYERS}

In the following, we rely on mean 3D values, hereafter $\langle {\rm
  3D}\rangle$, for all quantities except for the asymptotic entropy
$s_{\rm env}$. $\langle {\rm 3D}\rangle$ values are the temporal and
spatial average of 3D simulations over constant geometrical depth. We
use 250 snapshots in the last 25\% of a simulation to make the
temporal average. While our earlier studies have relied on averages
over constant optical depth, the geometrical depth is better suited to
extract convective fluxes and overshoot velocities.

Before comparing 3D simulations and 1D envelopes, it is crucial to
define what we refer to as the convection zone. In the local MLT
picture, the convective regions are clearly characterized as the
layers where the radiative gradient

\begin{equation}
\nabla_{\rm rad} = \left( \frac{\partial \ln T}{\partial \ln P}\right)_{\rm rad} ~,
\end{equation}

{\noindent}is larger than the adiabatic gradient

\begin{equation}
\nabla_{\rm ad} = \left( \frac{\partial \ln T}{\partial \ln P}\right)_{\rm ad} ~,
\end{equation}

{\noindent}with $T$ the temperature and $P$ the pressure. All other
parts of the structure are fully static. This is a rather crude
approximation of the dynamical nature of convection, where material
flows do not vanish abruptly when the thermal structure becomes
stable. In this section, we review the different regions that are
found in non-local models of the lower part of convection zones
\citep[see also][]{skaley91,chan92,freytag96}. Table~1 formally
defines the regions discussed in this section, and we give an example
of the geometrical extent and mass included in these layers based on
the 12,100~K and $\log g = 8$ simulation.

 \begin{deluxetable*}{lccccccccc}
 \tabletypesize{\scriptsize}
 \tablecolumns{8}
 \tablewidth{0pt}
 \tablecaption{Regions in the Lower Part of Convection Zones}
 \tablehead{
 \colhead{Region} &
 \colhead{$\frac{ds}{dz}$ {\tablenotemark{a}}} &
 \colhead{$\frac{ds}{dz}$ {\tablenotemark{a}}} &
 \colhead{$F_{\rm conv}/F_{\rm total}$} &
 \colhead{$F_{\rm conv}/F_{\rm total}$} &
 \colhead{$v_{z}$} &
 \colhead{$v_{z}$} &
 \colhead{$\Delta z$ {\tablenotemark{b}}} &
 \colhead{$\Delta \log M_{\rm H}/M_{\rm tot}$ {\tablenotemark{b}}} & \\
 \colhead{} &
 \colhead{(3D)} &
 \colhead{(1D)} &
 \colhead{(3D)} &
 \colhead{(1D)} &
 \colhead{(3D)} &
 \colhead{(1D)} &
 \colhead{} &
 \colhead{}
 }
 \startdata
Zone 1 & $< 0$ & $< 0$ & $>0$     &  $>0$ & $\neq 0$  & $\neq 0$ & $-$ & $-$ \\
Zone 2 & $> 0$ & $> 0$ & $>0$     &  $0$  & $\neq 0$  & $0$      & $0.8 H_{\rm P}$ & 0.2 \\
Zone 3 & $> 0$ & $> 0$ & $<0$     &  $0$  & $\neq 0$  & $0$      & $\sim 1.6 H_{\rm P}$ & $\sim0.5$ \\
Zone 4 & $> 0$ & $> 0$ & $\sim 0$ &  $0$  & $\neq 0$  & $0$      & $> 3 H_{\rm P}$ & $> 1.0$ \\
 \enddata

\tablenotetext{a}{The coordinate $z$ points towards the exterior of the star.}
\tablenotetext{b}{Ranges are taken from the simulation at 
$T_{\rm eff} = 12,100$ K and $\log g \sim 8.0$ as an illustrative
example. Zones 3 and 4 feature an exponential decay of (negative) flux
and velocity, respectively, and their depth can only be defined
approximately. For the example presented here, we adopt a bottom
boundary of $\left| F_{\rm conv}/F_{\rm total} \right| < 0.1$ for Zone
3.}
\label{tb:3D}
\end{deluxetable*}

\begin{figure}[h!]
\epsscale{0.8}
\plotone{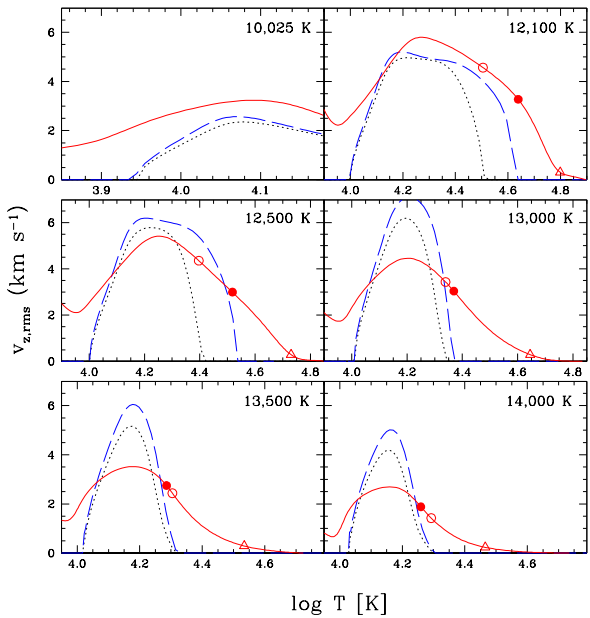}
\caption{Vertical RMS velocity as a function of the logarithm of the
  temperature for 3D simulations at $\log g = 8$ (solid red
  lines). The $T_{\rm eff}$ values are identified on the top right of
  the panels. We show the position of the Schwarzschild boundary
  (open circles), the flux boundary (filled circles), and the $v_{\rm
    z,rms}$ decay of 1 dex (open triangles) below the value at the
  flux boundary. We also display 1D model atmospheres with the
  calibration of the MLT parameters (see Table~2) for the 
    Schwarzschild (dotted black) and flux boundaries (dashed
  blue). For the models warmer than 13,000~K, we rely on an asymptotic
  parameterization of ML2/$\alpha_{\rm Schwa}$ = 1.2 and
  ML2/$\alpha_{\rm flux}$ = 1.4, respectively (see Sect. 5.1).
 \label{fg:f_vel}}
\end{figure}

To further illustrate the profile of 3D convection zones,
Figure~\ref{fg:f_vel} displays the RMS vertical velocities for
closed-box simulations at $\log g = 8$. We start from the vertical
velocity

\begin{equation}
v_{\rm z} = u_{\rm z} - \frac{\langle \rho
    u_{\rm z} \rangle}{\langle \rho \rangle}~,
\end{equation}

{\noindent}where the mass flux weighted mean velocity (second term on
right-hand side) is removed from the directly simulated velocity
  $u_{\rm z}$ to account for the residual numerical mass flux. The
latter results from the presence of plane-parallel oscillations and an
imperfect temporal averaging due to the finite number of
  snapshots. The corresponding RMS vertical velocity is
    
\begin{equation}
v^2_{\rm z,rms} = \langle v_{\rm z}^2\rangle = \langle u_{\rm z}^2 \rangle + \frac{\langle \rho
  u_{\rm z} \rangle^2}{\langle \rho \rangle^2} - 2\frac{\langle \rho
  u_{\rm z} \rangle \langle u_{\rm z} \rangle}{\langle \rho \rangle}~,
\end{equation}

\begin{figure}[h!]
\epsscale{0.8}
\plotone{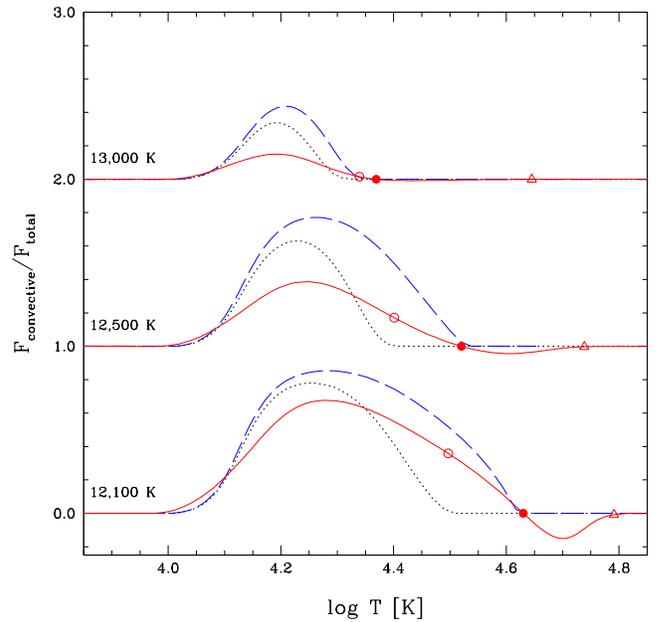}
\caption{Ratio of the convective energy flux to total flux as a
  function of the logarithm of the temperature at $\log g = 8$. The
  $\langle {\rm 3D}\rangle$ fluxes are represented by solid red lines
  and $T_{\rm eff}$ values for the simulations are identified on the
  panel. The ratio is exact for the 12,100~K model, but other
  structures are shifted by one flux units for clarity. The symbols
  are the same as in Figure~\ref{fg:f_vel}. We also display 1D model
  atmospheres matching the Schwarzschild boundary (black,
  dotted) and the flux boundary (blue, dashed). Parameters for the
  Schwarzschild boundary are ML2/$\alpha_{\rm Schwa}$ = 0.88,
  1.07, and 1.32, for the 12,100, 12,500, and 13,000 K models,
  respectively. The values are ML2/$\alpha_{\rm flux}$ = 1.00, 1.25,
  and 1.50 for the flux boundary at the same
  temperatures. \label{fg:f_flux}}
\end{figure}

\begin{figure}[h!]
\epsscale{0.8}
\plotone{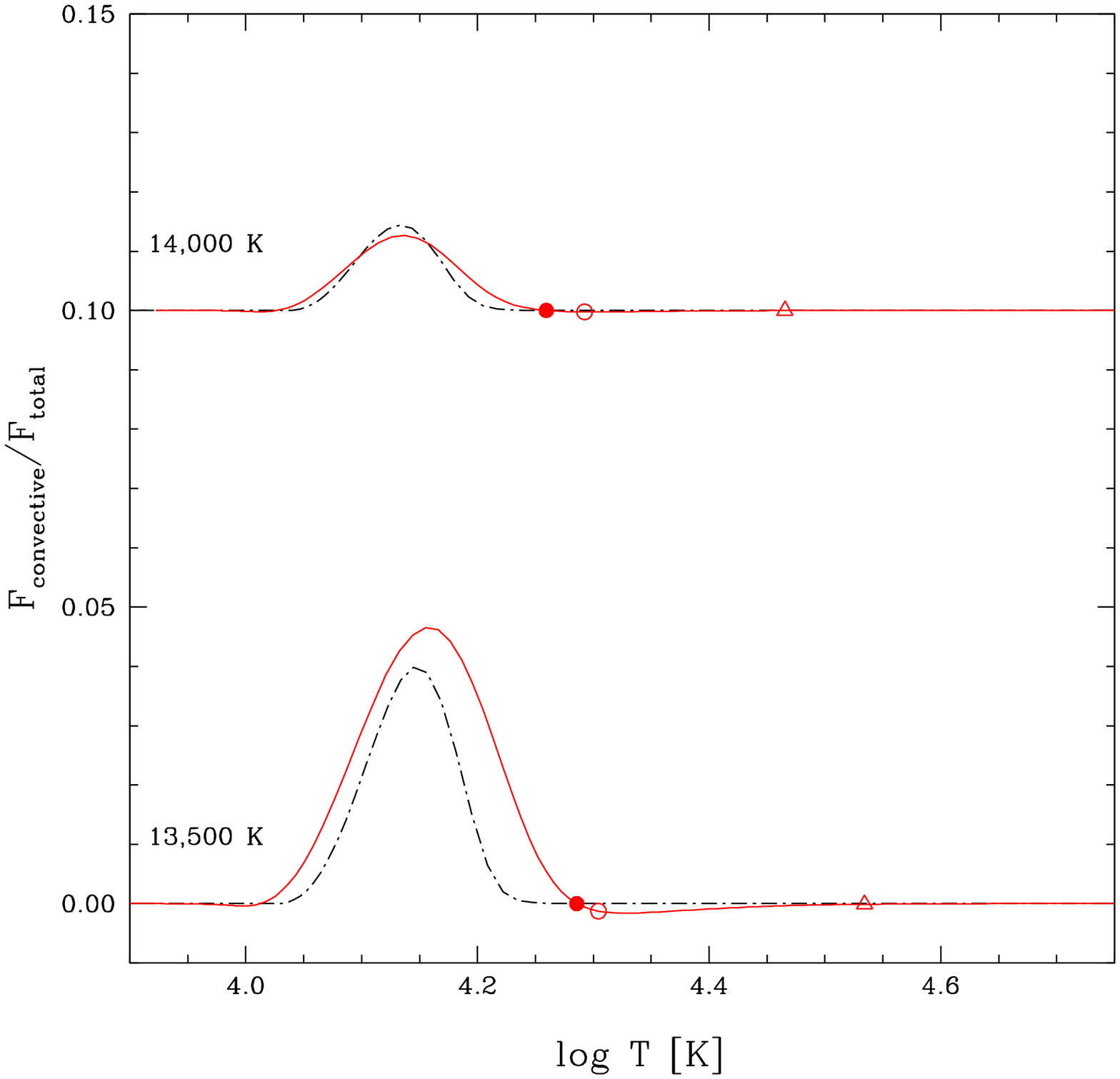}
\caption{Similar to Figure~\ref{fg:f_flux} but for the 3D simulations
  (solid red) at 13,500 and 14,000~K. The convective to total flux
  ratio is exact for the model at 13,500~K and shifted by 0.1 flux
  units for the 14,000~K case. The symbols are the same as in
  Figure~\ref{fg:f_vel}. In this regime, the MLT is unable to
  replicate both the 3D size of the convection zone and the maximum
  $F_{\rm conv}/F_{\rm tot}$ ratio. We display instead 1D ML2/$\alpha$
  = 0.7 model atmospheres (black, dot-dashed), which correspond to the
  average MLT parameterization to reproduce the maximum $F_{\rm
    conv}/F_{\rm tot}$ ratio for shallow convection zones (see
  Sect. 4.1).
\label{fg:f_flux2}}
\end{figure}

{\noindent}where all averages are performed over constant geometrical
depth\footnote{This differs from the RMS velocity fluctuation $\langle
  v_{\rm z}^2 \rangle - \langle v_{\rm z} \rangle^2$ where $\langle
  v_{\rm z} \rangle$ is expected to be non-zero due to a correlation
  between velocity and density fluctuations in the convection
  zone.}. Furthermore, Figures~\ref{fg:f_flux} and \ref{fg:f_flux2}
show the $\langle {\rm 3D}\rangle$ and 1D convective flux
profiles. The $\langle {\rm 3D}\rangle$ convective flux is the sum of
the enthalpy and kinetic energy fluxes,

\begin{equation}
F_{\rm conv} = \langle (e_{\rm int} + \frac{P}{\rho}){\rho u_{\rm z}} \rangle +
\langle \frac{{\mathbf u}^{2}}{2}\rho u_{\rm z} \rangle - e_{\rm tot} \langle \rho u_{\rm z}
\rangle ~,
\end{equation}

{\noindent}where $e_{\rm int}$ is the internal energy per gram, $\rho$
the density, and ${\mathbf u}$ the 3D velocity. The mass flux weighted energy
flux (third term on right-hand side of Eq.~(5)) is subtracted to
correct for any residual non-zero mass flux in the numerical
simulations as in Eq.~(4). This correction is a small fraction of the
convective flux for all simulations. The total energy is defined from

\begin{equation}
e_{\rm tot} = \frac{\langle \rho e_{\rm int} + P + \rho
  \frac{{\mathbf u}^2}{2}\rangle}{\langle \rho \rangle}~.
\end{equation}

{\noindent}We use the logarithm of the temperature as an independent
variable since it is a local quantity, while optical depth and mass
are integrated from the top of the convection zone, and are more
sensitive to differences in the photosphere.

The proper convection zone in 3D (open circles in
Figs.~\ref{fg:f_vel}-\ref{fg:f_flux2}) is defined in the same way as
in 1D from the Schwarzschild (stability) criterion. In this region,
the entropy gradient is negative with respect to geometrical depth
(increasing towards the exterior). In the following, we define the
bottom of this region as the {\it Schwarzschild boundary}. In the 3D
simulations, convective flows are largely created, horizontally
advected, and merged into narrow downdrafts in the photosphere
\citep{freytag96}. Large entropy fluctuations are produced by the
radiative cooling in these layers, which drives the convective
motions. For cool convective white dwarfs ($T_{\rm eff} \lesssim
11,000$ K, $\log g = 8$) with deep convection zones, entropy
  fluctuations are smaller in the photosphere and the dominant role of
  the downflows is diminished. The descending fluid form a
  hierarchical structure of merging downdrafts due to the increase of
  the pressure scale height with depth \citep{asplund09}.

In the 3D simulations, downdrafts at the base of the convection zone
(according to the Schwarzschild criterion) still have large
momenta. They are also denser than the ambient medium, albeit with a
decreasing difference. As a consequence, the convective cells are
still accelerated in the region just below the unstable layers. Mass
conservation guarantees that there is warm material transported
upwards, hence there is a positive convective flux is this
region. These layers are equivalent to a convection zone in
thermodynamical terms. We define the bottom of this region as the
layer where $F_{\rm conv}/F_{\rm tot} = 0$ and refer to it as the {\it
  flux boundary} (filled circles in
Figs.~\ref{fg:f_vel}-\ref{fg:f_flux2}). The typical size of the region
between the Schwarzschild and flux boundaries is a bit less than one
pressure scale height, or $\sim$0.3 dex in mass.

At the flux boundary, the momentum of the downdrafts remains
significant, hence they penetrate into even deeper layers. This is the
beginning of the convective overshoot region, although some authors
prefer the term convective penetration \citep{zahn91} when the
convective flux is still energetically relevant. In these layers,
convective structures are decelerated since they have a density
deficit. Downdrafts are generally warmer than the ambient medium and
they carry a net downwards (negative) convective flux, or in other
words, the temperature gradient in these layers is larger than the
radiative gradient. That follows from the change of sign of the
velocity-enthalpy correlation. However, Figure~\ref{fg:f_flux}
demonstrates that this negative overshoot flux is always a small
fraction ($\lesssim 10\%$) of the total flux. Once the convective flux
has decreased by one order of magnitude, or to a value of less than
1\% of the total flux, the energetic impact on the structure becomes
very small.

The negative convective flux and velocities decay in a similar
exponential way below the flux boundary, both with a scale height
close to $H_{\rm P}$. While the convective flux becomes rapidly
energetically negligible, the convective velocities still have mixing
capabilities in much deeper layers. This situation is due to the
extreme ratio between convective and diffusive time scales (see
Section~5.2). In typical cases for DA white dwarfs, convective
velocities are of the order of $v_{\rm z, rms} \sim 1$ km s$^{-1}$ at
base of the convection zone, while overshoot velocities of the order
of 1 m s$^{-1}$ still dominate over the slower diffusive speeds, and
can efficiently mix elements \citep{freytag96}. This implies that
microscopic diffusion timescales are likely to dominate only in the
deep overshoot layers, i.e. a few $H_{\rm P}$ below the flux
boundary. The exact layer where this happens depends on the diffusing
trace chemical element and the atmospheric parameters of the model,
although it is clear that the mixed region can be much larger than in
the 1D approximation. In Figures~\ref{fg:f_vel}-\ref{fg:f_flux2}, we
identify the position of a 1 dex velocity decay with respect to the
velocity at the flux boundary (filled triangles), which is generally
close to the bottom of the simulation. Our simulations evidently
provide a truncated picture of the overshoot layers and we review this
issue in Section~5.2.

\section{COMPARISON OF 1D AND 3D CONVECTION ZONES}

\subsection{Closed 3D Simulations}

\begin{figure}[h!]
\epsscale{0.8}
\plotone{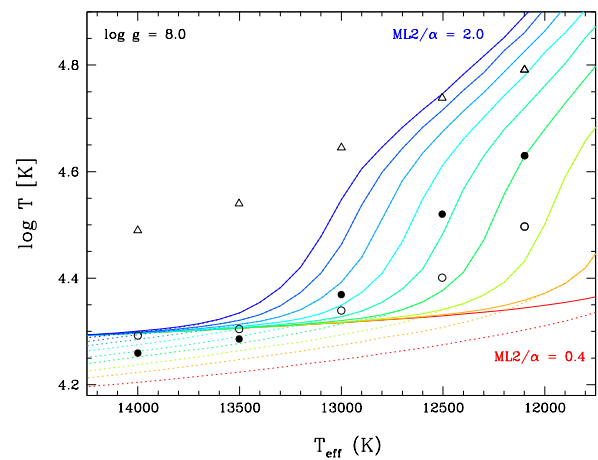}
\caption{Logarithm value of the temperature at the bottom of the
  convection zone as a function of $T_{\rm eff}$, for DA white dwarf
  envelopes at $\log g = 8$. The $\langle {\rm 3D}\rangle$ results are
  shown with black symbols using different definitions for the bottom
  of the convection zone (see Sect.~3). We consider the 
    Schwarzschild boundary (open circles), the flux boundary (filled
  circles), and a $v_{\rm z,rms}$ decay of 1 dex (open triangles)
  below the value at the flux boundary. We also display 1D sequences
  with the MLT parameterization varying from ML2/$\alpha$ = 0.4 (red)
  to 2.0 (blue) in steps of 0.2 dex. The solid lines represent the
  bottom of the convection zone defined by the Schwarzschild
  boundary while the dotted lines stand for the layers below which the
  convective flux becomes energetically negligible ($F_{\rm
    conv}/F_{\rm tot} < 0.01$). \label{fg:f_t_top}}
\end{figure}

\begin{figure}[h!]
\epsscale{0.8}
\plotone{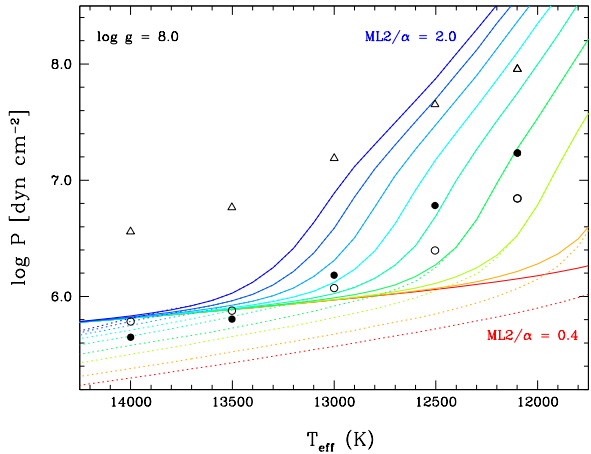}
\caption{Similar to Figure~\ref{fg:f_t_top} but for the thermal
  pressure (logarithm value) at the bottom of the convection zone as a
  function of $T_{\rm eff}$ at $\log g = 8$. \label{fg:f_p_top}}
\end{figure}

We first proceed with a direct comparison of $\langle {\rm 3D}\rangle$
and 1D stratifications in the case of shallow convection zones,
completely enclosed within the simulation
domain. Figures~\ref{fg:f_t_top} and \ref{fg:f_p_top} present the
$\langle {\rm 3D}\rangle$ logarithmic values of the temperature and
pressure, respectively, characterizing the bottom of the convection
zone for simulations at $\log g = 8$. We rely on three different
definitions for the size of the convection zone as discussed in
Section~3, with the same symbols as in
Figures~\ref{fg:f_vel}-\ref{fg:f_flux2}. These regions correspond to
the Schwarzschild boundary (open circles), the flux boundary (filled
circles), and a $v_{\rm z,rms}$ decay of 1 dex (open triangles) below
the reference value at the flux boundary. Figures~\ref{fg:f_t_top} and
\ref{fg:f_p_top} also display 1D sequences, with values ranging from
ML2$/\alpha$ = 0.4 to 2.0 in steps of 0.2, using the Schwarzschild
boundary to define the size of the convection zone (solid lines).

Figures~\ref{fg:f_t_top} and~\ref{fg:f_p_top} demonstrate that the 1D
envelope that best matches the bottom of a 3D simulation is generally
independent of whether the matching is performed on temperature
  or pressure. The pressure is proportional to the 1D mass column,
where only thermodynamic pressure contributes to hydrostatic
equilibrium, while in 3D simulations one must also account for the
turbulent pressure. Figure~\ref{fg:f_mass} depicts the $\langle {\rm
  3D}\rangle$ and 1D comparison in terms of the hydrogen mass, and the
results are similar to those presented for the temperature and
pressure at the base of the convection zone. It implies that even
though $\langle {\rm 3D}\rangle$ and 1D models have different profiles
in the photosphere, due to the top convective overshoot and turbulent
pressure, differences in the integrated mass column are small in the
lower part of the convection zone. In the following, the calibrated
ML2$/\alpha$ is the average value of the two 1D models that best match
the $\langle {\rm 3D}\rangle$ pressure and temperature at the bottom
of the convection zone, respectively, within a prescribed
boundary. The mass column can be directly extracted from the envelopes
calibrated in this way.

In terms of the Schwarzschild boundary, Figures~\ref{fg:f_t_top}
and~\ref{fg:f_p_top} demonstrate that the mixing-length parameter
increases rapidly with $T_{\rm eff}$, with values of 0.88, 1.07, and
1.32 at 12,100, 12,500, and 13,000~K, respectively. In this $T_{\rm
  eff}$ range partially covering the ZZ Ceti instability strip, the
MLT variation is significant compared to the usually assumed constant
value of ML2/$\alpha$ = 1.0 for envelopes \citep{fontaine08}. Our
  calibration of ML2/$\alpha$ is meant to represent the $\langle {\rm
  3D}\rangle$ temperature and pressure at the Schwarzschild boundary,
and by construction, it provides an estimation of the average
temperature gradient for the full convection zone. However, the
photospheric temperature gradient of a calibrated 1D envelope is not
expected to correspond to that of the 3D simulation.

\begin{figure}[h!]
\epsscale{0.8}
\plotone{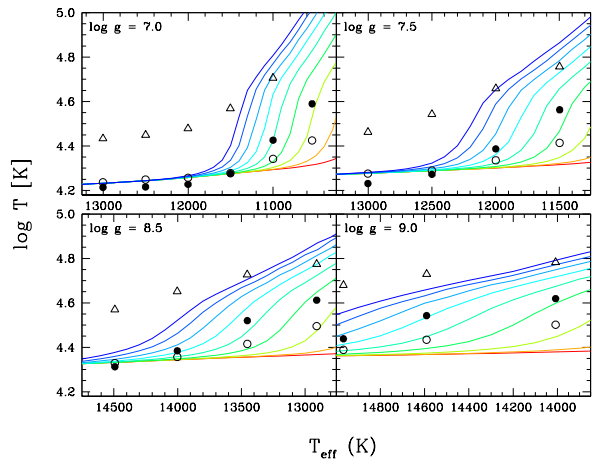}
\caption{Similar to Figure~\ref{fg:f_t_top} but for $\log g$ = 7.0,
  7.5, 8.5, and 9.0, with values identified on the
  panels. \label{fg:f_top_all}}
\end{figure}

For the convection zone defined in terms of the $\langle {\rm
  3D}\rangle$ flux boundary, ML2/$\alpha$ values have to be increased
to 1.00, 1.25, and 1.50 for the same $T_{\rm eff}$ values as
above. The derived efficiency is significantly higher than that found
for the Schwarzschild boundary. One should be cautious since an
inspection of Figure~\ref{fg:f_flux} for 1D model atmospheres
calibrated for the flux boundary (blue dashed lines) reveals that
while the zero point of convective flux is by definition in agreement
with the 3D simulations, the overall shape of the $\langle {\rm
  3D}\rangle$ convective flux is not very well reproduced for shallow
convection zones. {Our calibration of ML2/$\alpha$} is mostly useful
to characterize the depth at which convection becomes energetically
insignificant and the velocities start to decay exponentially with
geometrical depth. Finally, Figure~\ref{fg:f_top_all} demonstrates
that the $\langle {\rm 3D}\rangle$ versus 1D results (temperature
only) at other gravities are fairly similar, albeit with a shift in
$T_{\rm eff}$. As a consequence, the previous discussion applies most
generally to white dwarfs with shallow convection zones.

For the very warm simulations, e.g. 13,500 and 14,000~K at $\log g$ =
8.0, the Schwarzschild and flux boundaries are essentially in the
photosphere ($\tau_{\rm R, bottom} < 10$), and therefore the
ML2/$\alpha$ value for these layers becomes coupled with the MLT
parameterization used in spectroscopic applications (TL13c). Both 3D
simulations and 1D models show new patterns in this $T_{\rm eff}$
regime. The $\langle {\rm 3D}\rangle$ convective flux becomes
negligible outside of the unstable layers, and there is a reversal of
the flux and Schwarzschild boundaries, with the Schwarzschild boundary
moving below the flux boundary with increasing $T_{\rm eff}$. In this
regime, efficient radiation transport is able to smooth temperature
fluctuations. This diminishes the flux of internal energy (first term
in Eq.~(5)) over a shorter distance from the top of the convection
zone than the velocity field becomes symmetric in up- and downflows.
The significant momentum of the narrow downdrafts produces a negative
kinetic energy flux (second term in Eq.~(5)). This flux remains large
near the mean Schwarzschild boundary since cool downdrafts get
convectively stable at larger geometrical depths than the upflows. As
a consequence, the mean total flux becomes negative slightly above the
mean Schwarzschild boundary. We have verified that there is no
reversal of the flux and Schwarzschild boundaries when the kinetic
energy flux is neglected.  The MLT does not account for the kinetic
energy flux, hence we do not expect a similar reversal in 1D.

For convective 1D models at large $T_{\rm eff}$, the size of the
unstable regions becomes insensitive to the MLT parameterization
according to Figure~\ref{fg:f_t_top}, hence it is not possible to
calibrate the MLT based on the Schwarzschild boundary. This picture is
somewhat misleading since the MLT convective fluxes, and associated
velocities, remain very sensitive to the value of the MLT
parameters. Figure~\ref{fg:f_t_top} shows that the 1D convective flux
drops to very small values ($F_{\rm conv}/F_{\rm tot} < 0.01$, dotted
lines) much higher in the photosphere than the 1D Schwarzschild
boundary.

\begin{figure}[h!]
\epsscale{0.8}
\plotone{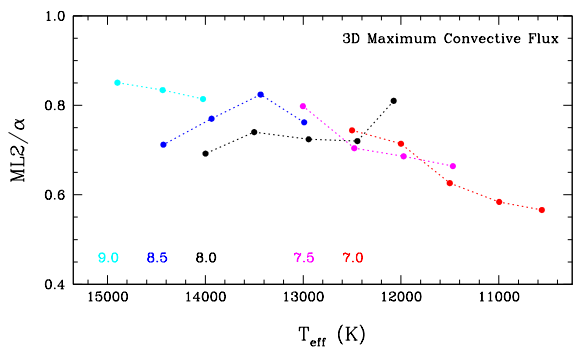}
\caption{Calibration of ML2/$\alpha$ for the maximum $F_{\rm
    conv}/F_{\rm tot}$ ratio as a function of $T_{\rm eff}$ and $\log
  g$ (represented by different colors with the legend at the
  bottom). The calibration is based on the 1D model that best
  replicates the maximum $\langle {\rm 3D}\rangle$ convective flux of
  closed simulations. This calibration can not be performed for deep
  convection zones since all 1D models have $F_{\rm conv,max}/F_{\rm
    tot} \sim$ 1.
   \label{fg:f_fmax}}
\end{figure}

Our results would naively suggest that convective efficiency increases
with $T_{\rm eff}$ but the 3D simulations present a more complex
picture. At high $T_{\rm eff}$, non-local effects from strong and deep
reaching downdrafts create $\langle {\rm 3D}\rangle$ flux profiles
that are extended and smoother as a function of geometrical depth than
in the 1D case, both at the top and bottom of the convection zone. In
Figure~\ref{fg:f_fmax}, we have calibrated ML2/$\alpha$ in order to
reproduce the maximum value of the $\langle {\rm 3D}\rangle$
convective flux, which peaks in the photosphere, for shallow
convection zones. Clearly, a much smaller mixing-length parameter is
necessary to match the $\langle {\rm 3D}\rangle$ convective flux in
the photosphere in comparison to the Schwarzschild or flux
boundaries. The values of ML2/$\alpha$ = 0.6-0.8 are consistent with
the commonly used spectroscopic parameterizations (TL13c).
Nevertheless, the parameterizations for the Schwarzschild and flux
boundaries offer a better representation of the conditions at the
bottom of the convection zones.

We have already discussed the fact that the convection zone is
drastically deeper when defined in terms of the $\langle {\rm
  3D}\rangle$ convective velocities.  This is also seen in
Figures~\ref{fg:f_t_top} and \ref{fg:f_p_top} where we show the
position of the one order of magnitude decrease for $v_{\rm z,rms}$
below the flux boundary (open triangles).  It is inappropriate to
parameterize the 1D MLT for the highly non-local overshoot velocities,
and it would produce spurious stratifications in the unstable regions.
Instead, we propose an overshoot parameterization that does not
directly involve the MLT in Section~5.2.

\subsection{Open 3D Simulations}

For open 3D simulations, we have extracted the asymptotic entropy
values $s_{\rm env}$ characterizing the deep adiabatic layers using
the technique described in Section~2.2. $s_{\rm env}$ is directly
derived from the specified entropy of the ascending material at the
bottom boundary of the simulations. We have verified that this matches
the observed asymptotic value below the photosphere (see, e.g.,
Figure~\ref{fg:f_exemple}). We then assume that $s_{\rm env}$ also
corresponds to the entropy value at the bottom of the unstable layers
in 1D envelopes. The $s_{\rm env}$ and 1D entropy values are compared
in Figure~\ref{fg:f_entropy} for the $\log g = 8$ case. The
calibration of ML2/$\alpha$ is directly performed from a match of
$s_{\rm env}$ with entropy values interpolated from the grid of 1D
envelopes. In Figure~\ref{fg:f_mass}, we show the resulting hydrogen
mass integrated from the surface.

At low temperatures ($T_{\rm eff} \lesssim 7000$~K at $\log g = 8$),
DA white dwarfs have extremely small super-adiabatic atmospheric
layers, and the structure remains essentially adiabatic from the
bottom to the top of the convection zone. Since the top of the
convection zone is higher than the photosphere ($\tau_{\rm R} \sim
0.1$), the effective temperature directly identifies the entropy value
at the bottom of the convection zone. The choice of the MLT
parameterization does not matter since there is no significant
radiative energy exchange during one advective (turnover) timescale.

\section{DISCUSSION}

\subsection{1D MLT Calibration}

\begin{figure}[h!]
\epsscale{0.8}
\plotone{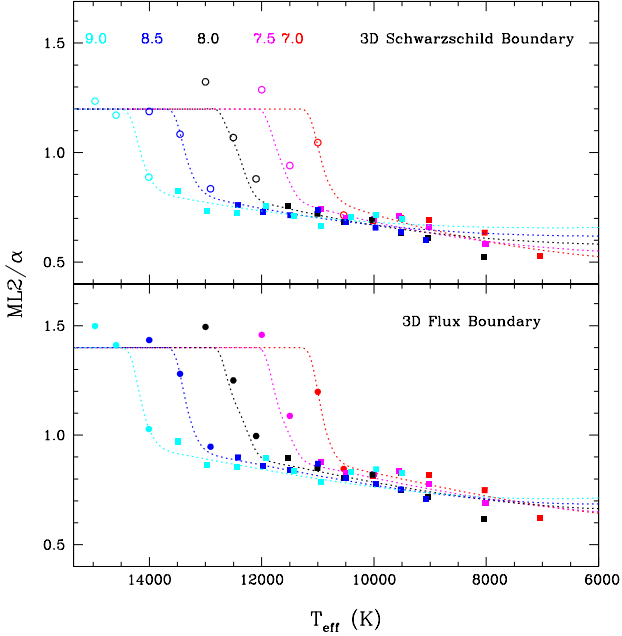}
\caption{{\it Top:} Calibration of ML2/$\alpha_{\rm Schwa}$ for the
  lower part of the convection zone as a function of $T_{\rm eff}$ and
  $\log g$ (represented by different colors with the legend at the
  top). The calibration is based on the 1D model that best replicates
  the Schwarzschild boundary of a 3D simulation, either from a
  direct comparison (open circles) or by using the $s_{\rm env}$
  calibration (filled squares). The dotted lines correspond to the
  proposed fitting function (Eq.~(9)). {\it Bottom:} Calibration of
  ML2/$\alpha_{\rm flux}$ based on the 1D model that best represents
  the flux boundary of a 3D simulation (filled circles).  For open 3D
  simulations, we use ML2/$\alpha_{\rm flux}$ = 1.16 ML2/$\alpha_{\rm
    Schwa}$. The dotted lines correspond to the proposed fitting
  function (Eq.~(10)).
   \label{fg:f_xmlt}}
\end{figure}

Figure~\ref{fg:f_xmlt} (top panel) presents the MLT parameterization
for the lower part of the convection zone in order to recover the
Schwarzschild boundary (hereafter ML2/$\alpha_{\rm Schwa}$) of the 70
3D simulations in our grid. We illustrate with different symbols the
calibration derived directly from closed 3D simulations (open circles)
and inferred from a match of $s_{\rm env}$ (filled squares). We also
present in Figure~\ref{fg:f_xmlt} (bottom panel) the calibration
matching the $\langle {\rm 3D}\rangle$ flux boundary (hereafter
ML2/$\alpha_{\rm flux}$). The latter calibration is directly performed
for closed simulations, and in those cases, $\alpha_{\rm flux}$ is
16\% larger than $\alpha_{\rm Schwa}$ with a relatively small
dispersion of 3\%. Therefore, we simply assume that ML2/$\alpha_{\rm
  flux} = 1.16~{\rm ML2/}\alpha_{\rm Schwa}$ for open 3D
simulations. This is likely a good approximation in the transition
region between closed and open 3D simulations, and at lower $T_{\rm
  eff}$, the 1D envelopes depend less critically on the MLT
parameterization.

The calibration is not performed when the 1D mass included within the
convection zone varies by an amount smaller than 0.2 dex for the range
of ML2$/\alpha$ between 0.4 and 2.0. This defines the upper and lower
$T_{\rm eff}$ boundaries in Figure~\ref{fg:f_xmlt}, which depend on
$\log g$. At the cool end, we propose to keep ML2/$\alpha$ constant,
since the value is irrelevant for structure calculations. Similarly,
at $T_{\rm eff}$ values above those in the calibration range, it is
likely acceptable to keep the value constant for most
applications. The choice of the asymptotic ML2/$\alpha$ value is not
obvious, however, because of its rapid variation with $T_{\rm
  eff}$. As a compromise, we adopt values of 1.2 and 1.4, for
ML2/$\alpha_{\rm Schwa}$ and ML2/$\alpha_{\rm flux}$, respectively, at
$T_{\rm eff}$ values larger than our calibration range. If one is
interested in the detailed properties of shallow convection zones
above $T_{\rm eff} \sim 12,000$~K at $\log g$ = 8.0, it may be
preferable to combine the $\langle {\rm 3D}\rangle$ and 1D structures
at some depth below the convection zone where the convective flux is
negligible. The MLT does not reproduce very well the extended but
inefficient 3D convection zones in this regime.  For $T_{\rm eff}$
values above our calibration range, most of the 3D effects will be
from the overshoot at the base of the convection zone since contrary
to the small convective fluxes, velocities remain significant well
below the photosphere.

Table~2 provides the tabulated MLT parameterizations, which are valid
for 1D envelopes with an EOS, opacities, and boundary conditions
similar to those employed for our grid. Physical conditions at the
bottom of our calibrated envelopes (mass, temperature, and pressure)
are also given as a reference point. Moreover, we propose fitting
functions for ML2/$\alpha_{\rm Schwa}$ and ML2/$\alpha_{\rm flux}$,
respectively, where the independent variables are defined as
  
\begin{equation}
g_{0} = \log g {\rm [cgs]} - 8.0 ~,
\end{equation}

\begin{equation}
T_{0} = (T_{\rm eff} {\rm [K]} -12000)/1000-1.6g_{0} ~,
\end{equation}

{\noindent}and the functions are as follow with numerical coefficients
found in Table~3

\begin{eqnarray}
{\rm ML2/}\alpha_{\rm Schwa} = 
\Big(a_{1}+(a_{2}+a_{3}\exp[a_{4}T_{0}+a_{5}g_{0}]) \nonumber \\
\exp\Big[(a_{6}+a_{7}\exp[a_{8}T_{0}])T_{0}+a_{9}g_{0}\Big]\Big) \nonumber \\
+a_{10}\exp\Big(-a_{11}([T_{0}-a_{12}]^2+[g_{0}-a_{13}]^2)\Big) ~,
\end{eqnarray}

\begin{eqnarray}
{\rm ML2/}\alpha_{\rm flux} =
\Big(a_{1}+a_{2}\exp\Big[(a_{3}+\{a_{4}+a_{5} \nonumber \\
\exp[a_{6}T_{0}+a_{7}g_{0}]\}\exp[a_{8}T_{0}])T_{0}+a_{9}g_{0}\Big]\Big) \nonumber \\
+a_{10}\exp\Big(-a_{11}([T_{0}-a_{12}]^2+[g_{0}-a_{13}]^2)\Big) ~.
\end{eqnarray}

{\noindent}The proposed functions are presented in
Figure~\ref{fg:f_xmlt} along with the data points.  Similarly to our
3D atmospheric parameter corrections in TL13c, we have adopted
functions that do not retain the fine details of the 3D and 1D
differences. Small scale fluctuations may be due to inaccuracies in
the grid of 3D simulations. Ultimately, the calibrated 1D structures
do not provide the detailed $\langle {\rm 3D}\rangle$ convective flux
profile and neglect the turbulent nature of convection. It is not well
constrained how much these 3D effects impact chemical diffusion and
pulsation calculations. Finally, we remind the reader that 1D
structure codes typically make approximations for the non-gray
radiative transfer in the atmospheres, which may introduce a slight
offset in the size of convection zones. The ML2/$\alpha$ offset is at
most a few percent for our setup (see Section 2.3). As a consequence,
we believe that a calibration within 5\% is sufficient.

 \begin{deluxetable*}{llllllllll}
 \tabletypesize{\scriptsize}
 \tablecolumns{10}
 \tablewidth{0pt}
 \tablecaption{ML2/$\alpha$ Calibration for DA Envelopes}
 \tablehead{
 \colhead{$T_{\rm eff}$} & 
 \colhead{$\log g$} & 
 \colhead{ML2/$\alpha_{\rm Schwa}$ {\tablenotemark{a}}} & 
 \colhead{$\log M_{\rm H}/M_{\rm tot}$ {\tablenotemark{a}}} & 
 \colhead{$\log T$ {\tablenotemark{a}}} & 
 \colhead{$\log P$ {\tablenotemark{a}}} & 
 \colhead{ML2/$\alpha_{\rm flux}$ {\tablenotemark{b}}} & 
 \colhead{$\log M_{\rm H}/M_{\rm tot}$ {\tablenotemark{b}}} & 
 \colhead{$\log T_{\rm}$ {\tablenotemark{b}}} & 
 \colhead{$\log P_{\rm}$ {\tablenotemark{b}}} \\
 \colhead{(K)} &
 \colhead{} &
 \colhead{} &
 \colhead{} &
 \colhead{[K]} &
 \colhead{[dyn cm$^{-2}$]} &
 \colhead{} &
 \colhead{} &
 \colhead{[K]} &
 \colhead{[dyn cm$^{-2}$]} 
 }
 \startdata
   6112& 7.00&  0.53& $-$6.06&  5.93& 14.04&  0.61& $-$6.03&  5.94& 14.07\\
   7046& 7.00&  0.53& $-$6.79&  5.80& 13.30&  0.61& $-$6.74&  5.81& 13.36\\
   8027& 7.00&  0.63& $-$7.51&  5.68& 12.58&  0.74& $-$7.42&  5.69& 12.67\\
   9025& 7.00&  0.69& $-$8.90&  5.41& 11.19&  0.80& $-$8.69&  5.45& 11.40\\
   9521& 7.00&  0.70&$-$10.14&  5.17&  9.94&  0.81& $-$9.78&  5.24& 10.31\\
  10018& 7.00&  0.69&$-$12.00&  4.84&  8.07&  0.80&$-$11.49&  4.93&  8.59\\
  10540& 7.00&  0.72&$-$13.92&  4.44&  6.16&  0.85&$-$13.33&  4.58&  6.74\\
  11000& 7.00&  1.05&$-$14.30&  4.34&  5.78&  1.20&$-$13.98&  4.42&  6.10\\
  11501& 7.00&  1.20&$-$14.63&  4.28&  5.45&  1.40&$-$14.59&  4.29&  5.49\\
  12001& 7.00&  1.20&$-$14.78&  4.26&  5.30&  1.40&$-$14.77&  4.26&  5.30\\
  12501& 7.00&  1.20&$-$14.89&  4.24&  5.19&  1.40&$-$14.89&  4.24&  5.19\\
  13003& 7.00&  1.20&$-$14.98&  4.23&  5.10&  1.40&$-$14.98&  4.23&  5.10\\
   6065& 7.50&  0.58& $-$6.91&  5.90& 14.17&  0.67& $-$6.90&  5.90& 14.19\\
   7033& 7.50&  0.58& $-$7.54&  5.79& 13.54&  0.67& $-$7.51&  5.80& 13.57\\
   8017& 7.50&  0.58& $-$8.22&  5.68& 12.86&  0.68& $-$8.16&  5.69& 12.92\\
   9015& 7.50&  0.66& $-$9.07&  5.53& 12.01&  0.77& $-$8.95&  5.55& 12.13\\
   9549& 7.50&  0.71& $-$9.83&  5.38& 11.25&  0.82& $-$9.64&  5.42& 11.44\\
  10007& 7.50&  0.70&$-$10.81&  5.19& 10.27&  0.81&$-$10.52&  5.25& 10.55\\
  10500& 7.50&  0.70&$-$12.30&  4.93&  8.78&  0.82&$-$11.86&  5.00&  9.21\\
  10938& 7.50&  0.74&$-$13.68&  4.68&  7.40&  0.86&$-$13.12&  4.79&  7.96\\
  11498& 7.50&  0.94&$-$14.79&  4.42&  6.29&  1.09&$-$14.27&  4.56&  6.81\\
  11999& 7.50&  1.20&$-$15.24&  4.32&  5.84&  1.40&$-$15.06&  4.36&  6.02\\
  12500& 7.50&  1.20&$-$15.43&  4.29&  5.65&  1.40&$-$15.42&  4.29&  5.66\\
  13002& 7.50&  1.20&$-$15.53&  4.28&  5.55&  1.40&$-$15.52&  4.28&  5.55\\
   5997& 8.00&  0.52& $-$7.68&  5.89& 14.40&  0.60& $-$7.67&  5.89& 14.41\\
   7011& 8.00&  0.52& $-$8.44&  5.75& 13.64&  0.60& $-$8.42&  5.76& 13.65\\
   8034& 8.00&  0.52& $-$9.01&  5.66& 13.07&  0.60& $-$8.97&  5.67& 13.11\\
   9036& 8.00&  0.61& $-$9.66&  5.55& 12.41&  0.71& $-$9.58&  5.57& 12.50\\
   9518& 8.00&  0.64&$-$10.10&  5.47& 11.98&  0.74& $-$9.98&  5.50& 12.10\\
  10025& 8.00&  0.69&$-$10.71&  5.36& 11.36&  0.80&$-$10.56&  5.39& 11.52\\
  10532& 8.00&  0.68&$-$11.61&  5.19& 10.46&  0.79&$-$11.39&  5.23& 10.68\\
  11005& 8.00&  0.72&$-$12.63&  5.00&  9.45&  0.84&$-$12.32&  5.06&  9.75\\
  11529& 8.00&  0.76&$-$14.04&  4.76&  8.04&  0.88&$-$13.58&  4.84&  8.50\\
  12099& 8.00&  0.88&$-$15.27&  4.51&  6.81&  1.00&$-$14.82&  4.63&  7.26\\
  12504& 8.00&  1.07&$-$15.69&  4.40&  6.38&  1.25&$-$15.27&  4.52&  6.81\\
  13000& 8.00&  1.20&$-$16.05&  4.33&  6.03&  1.40&$-$15.96&  4.35&  6.12\\
  13502& 8.00&  1.20&$-$16.18&  4.31&  5.89&  1.40&$-$16.17&  4.31&  5.90\\
  14000& 8.00&  1.20&$-$16.26&  4.30&  5.81&  1.40&$-$16.26&  4.30&  5.82\\
   6024& 8.50&  0.60& $-$8.51&  5.86& 14.57&  0.70& $-$8.51&  5.86& 14.57\\
   6925& 8.50&  0.60& $-$9.32&  5.71& 13.76&  0.70& $-$9.31&  5.72& 13.76\\
   8004& 8.50&  0.60& $-$9.80&  5.65& 13.28&  0.70& $-$9.78&  5.65& 13.30\\
   9068& 8.50&  0.60&$-$10.38&  5.55& 12.69&  0.70&$-$10.33&  5.56& 12.74\\
   9522& 8.50&  0.64&$-$10.67&  5.51& 12.41&  0.74&$-$10.61&  5.52& 12.47\\
   9972& 8.50&  0.66&$-$11.01&  5.45& 12.07&  0.76&$-$10.93&  5.46& 12.14\\
  10496& 8.50&  0.68&$-$11.52&  5.35& 11.55&  0.79&$-$11.41&  5.38& 11.67\\
  10997& 8.50&  0.74&$-$12.19&  5.23& 10.89&  0.86&$-$12.04&  5.26& 11.04\\
  11490& 8.50&  0.72&$-$13.04&  5.08& 10.04&  0.84&$-$12.81&  5.12& 10.26\\
  11979& 8.50&  0.73&$-$14.09&  4.89&  8.99&  0.84&$-$13.78&  4.95&  9.29\\
  12420& 8.50&  0.76&$-$15.11&  4.72&  7.96&  0.88&$-$14.68&  4.79&  8.40\\
  12909& 8.50&  0.84&$-$16.11&  4.50&  6.96&  0.95&$-$15.71&  4.61&  7.37\\
  13453& 8.50&  1.08&$-$16.47&  4.42&  6.61&  1.28&$-$16.10&  4.51&  6.97\\
  14002& 8.50&  1.19&$-$16.76&  4.36&  6.31&  1.40&$-$16.67&  4.38&  6.41\\
  14492& 8.50&  1.20&$-$16.89&  4.33&  6.18&  1.40&$-$16.87&  4.34&  6.20\\
   6028& 9.00&  0.70& $-$9.38&  5.80& 14.70&  0.81& $-$9.38&  5.80& 14.70\\
   6960& 9.00&  0.70&$-$10.24&  5.67& 13.84&  0.81&$-$10.23&  5.67& 13.84\\
   8041& 9.00&  0.70&$-$10.75&  5.59& 13.33&  0.81&$-$10.74&  5.60& 13.34\\
   8999& 9.00&  0.70&$-$11.10&  5.55& 12.98&  0.81&$-$11.09&  5.55& 12.99\\
   9507& 9.00&  0.71&$-$11.34&  5.51& 12.74&  0.82&$-$11.31&  5.52& 12.77\\
   9962& 9.00&  0.72&$-$11.59&  5.47& 12.48&  0.84&$-$11.56&  5.48& 12.52\\
  10403& 9.00&  0.71&$-$11.89&  5.42& 12.18&  0.82&$-$11.85&  5.43& 12.23\\
  10948& 9.00&  0.67&$-$12.37&  5.34& 11.70&  0.77&$-$12.30&  5.35& 11.78\\
  11415& 9.00&  0.71&$-$12.84&  5.25& 11.23&  0.83&$-$12.73&  5.27& 11.35\\
  11915& 9.00&  0.76&$-$13.47&  5.14& 10.61&  0.88&$-$13.33&  5.16& 10.75\\
  12436& 9.00&  0.72&$-$14.32&  4.99&  9.76&  0.84&$-$14.11&  5.03&  9.97\\
  12969& 9.00&  0.73&$-$15.29&  4.83&  8.79&  0.85&$-$15.02&  4.87&  9.06\\
  13496& 9.00&  0.82&$-$16.05&  4.70&  8.03&  0.96&$-$15.75&  4.75&  8.32\\
  14008& 9.00&  0.89&$-$16.92&  4.51&  7.15&  1.03&$-$16.52&  4.62&  7.56\\
  14591& 9.00&  1.17&$-$17.23&  4.44&  6.85&  1.40&$-$16.87&  4.53&  7.20\\
  14967& 9.00&  1.20&$-$17.47&  4.38&  6.61&  1.40&$-$17.34&  4.41&  6.73\\
 \enddata

\tablenotetext{a}{Corresponds to the position of the $\langle {\rm 3D}\rangle$ Schwarzschild boundary for closed simulations (see Section 4.1). For open simulations, the calibration is performed by matching the 3D $s_{\rm env}$ value with the 1D entropy at the bottom of the convection zone (see Section 4.2).}
\tablenotetext{b}{Corresponds to the position of the $\langle {\rm 3D}\rangle$ flux boundary for closed simulations. For open simulations, we simply assume that ML2/$\alpha_{\rm flux} = 1.16~{\rm ML2}/\alpha_{\rm Schwa}$ (see Section 5.1).}
\tablecomments{$T_{\rm eff}$ is the spatial and temporal average of the emergent flux. The RMS $T_{\rm eff}$ variations are found in Table~1 of TL13b. $\log M_{\rm H}/M_{\rm tot}$, $\log T$, and $\log P$ are extracted at the bottom of the convection zone from calibrated 1D envelopes. \label{tb:t2}}
\end{deluxetable*}

 \begin{deluxetable}{lll}
 \tabletypesize{\scriptsize}
 \tablecolumns{3}
 \tablewidth{0pt}
 \tablecaption{Coefficients for Fitting Functions}
 \tablehead{
 \colhead{Coefficient} & 
 \colhead{ML2/$\alpha_{\rm Schwa}$} & 
 \colhead{ML2/$\alpha_{\rm flux}$} 
 }
 \startdata
a$_{1}$ &    1.1989083E+00   &    1.4000539E+00 \\
a$_{2}$ & $-$1.8659403E+00   & $-$5.1134694E$-$01\\
a$_{3}$ &    1.4425660E+00   & $-$1.1159288E+00\\
a$_{4}$ &    6.4742170E$-$02 &    1.0083984E+00\\
a$_{5}$ & $-$2.9996192E$-$02 & $-$5.7427026E$-$02\\
a$_{6}$ &    6.0750771E$-$02 &    5.4884977E+00\\
a$_{7}$ & $-$5.2572772E$-$02 & $-$1.6106825E$-$02\\
a$_{8}$ &    5.4690218E+00   & $-$7.5656008E$-$03\\
a$_{9}$ & $-$1.6330177E$-$01 & $-$6.8772823E$-$02\\
a$_{10}$ &   2.8348941E$-$01 &    2.9166886E$-$01\\
a$_{11}$ &   1.7353691E+01   &    1.8977236E+01\\
a$_{12}$ &   4.3545950E$-$01 &    3.6544167E$-$01\\
a$_{13}$ &$-$2.1739157E$-$01 & $-$2.2859657E$-$01\\
 \enddata
\label{tb:t4}
\end{deluxetable}

\subsection{Parameterization of Overshoot Velocities}

We have so far neglected the convective overshoot below the flux
boundary. In most cases, the quantity of interest is the overshoot
velocity, which does not exist in the local MLT. In the following, we
aim at providing a parameterization for overshoot in regions below the
1D convection zone.

The spatial scales and timescales involved in convection and
microscopic diffusion differ by many orders of magnitude in typical
white dwarfs. It is therefore not possible for multi-dimensional
simulations to model both effects simultaneously. Instead, we depict
the far overshoot regime as a random walk process characterized by a
macroscopic diffusion coefficient, which simply counter-balances the
microscopic diffusion coefficient in 1D calculations. The mixed
regions are those where macroscopic diffusion dominates over
microscopic diffusion. \citet{freytag96} studied this random walk
process with tracer particles in 2D RHD simulations. They found that
the particles are immediately mixed within the convection zone, but
that the RMS vertical spread $\delta z_{\rm overshoot}$ in the
overshoot layers could be described from

\begin{equation}
\delta z_{\rm overshoot}^2 = 2 D_{\rm overshoot}(z)t~,
\end{equation}

{\noindent}where $D_{\rm overshoot}$ is the macroscopic diffusion
coefficient

\begin{equation}
D_{\rm overshoot}(z) = v_{\rm z,rms}^2(z) t_{\rm char}(z)~,
\end{equation}

{\noindent}with $t_{\rm char}$ a characteristic timescale. Just based
on MLT models or even with detailed RHD simulations, $v_{\rm z,rms}$
is not directly available for the deep overshoot regions of
interest. As a consequence, \citet{freytag96} propose, from a match to
2D simulations and physical considerations, that $v_{\rm z,rms}$ has
an exponential decay below the convection zone.  The resulting
diffusion coefficient then takes the form

\begin{equation}
D_{\rm overshoot}(z) = v_{\rm base}^2 t_{\rm char} \exp(2(z-z_{\rm base})/H_{\rm v})~,
\end{equation}

{\noindent}where $v_{\rm base}$ is the velocity at the base of the
convection zone and $H_{\rm v}$ the velocity scale height. In the
following, we assume that the base of the convection zone is the flux
boundary as determined by 3D simulations and the 1D ML2$/\alpha_{\rm
  flux}$ parameterization.

\subsubsection{Closed 3D Simulations}

\begin{figure}[h!]
\epsscale{0.8}
\plotone{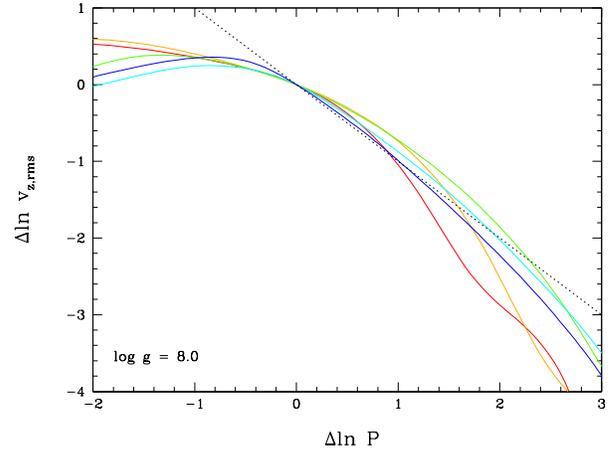}
\caption{Vertical RMS velocity decay as a function of pressure
  (natural logarithm values) for 3D simulations at $\log g = 8$. The
  reference point is the flux boundary for which we define
  $\Delta \ln v_{\rm z, rms} = 0$ and $\Delta \ln P = 0$. The
  simulations are color-coded from $T_{\rm eff}$ = 12,100 (red),
  12,500, 13,000, 13,500, to 14,000~K (blue). The $-1$ dotted black
  slope represents an exponential velocity decay with a scale height
  of $H_{\rm P}$. The velocity decay at $\Delta \ln P > 2$ could be
  impacted by the closed bottom boundary condition. \label{fg:f_hv}}
\end{figure}

For closed 3D simulations, it is possible to verify the proposed
exponential decay of overshoot velocities, as well as calibrate
Eq.~(13) by extracting $v_{\rm base}$, $t_{\rm char}$, and $H_{\rm
  v}$.  Figure~\ref{fg:f_hv} demonstrates that over the three pressure
scale heights typically included in our simulations below the flux
boundary, the velocity decay is nearly exponential. The velocity scale
height is very close to one pressure scale height (dotted black line),
although it is actually changing with depth. It is larger than one
pressure scale height immediately below the flux boundary, and becomes
subsequently smaller. As a consequence, taking $H_{\rm v} = H_{\rm P}$
is very likely to overestimate macroscopic diffusion in the deep
overshoot layers, and gives an upper limit to the mixed mass. Finally,
\citet{freytag96} demonstrate that the timescale of overshoot for
shallow convection zones is the same as the characteristic convective
timescale in the photosphere, since this is where the downdrafts are
formed. As a consequence, it is possible to use directly the
characteristic granulation timescales computed in TL13b and TL13c. In
Table~4, we present $v_{\rm z,rms}$ at the flux boundary ($v_{\rm
  base}$) and the characteristic granulation timescales ($t_{\rm
  char}$) for closed simulations, which can be used in Eq.~(13) for
shallow convection zones.  The velocity scale height can be directly
evaluated from the 1D pressure scale height in the envelopes since
this quantity is not significantly impacted by 3D effects, although we
also include the local $\langle {\rm 3D}\rangle$ values at the base of
the convection zone in Table~4.

\begin{figure}[h!]
\epsscale{0.8}
\plotone{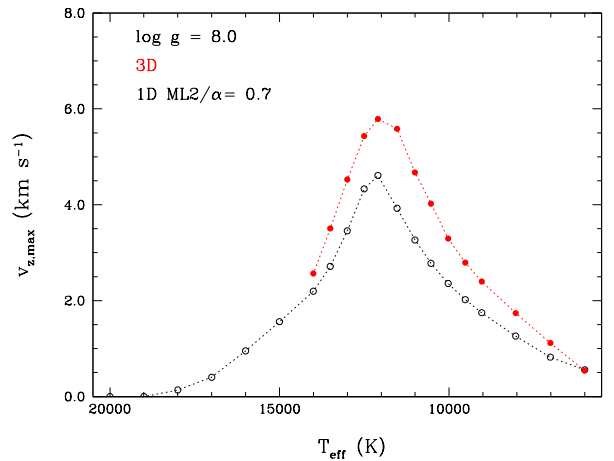}
\caption{Maximum $v_{\rm z,rms}$ velocity within the convection zone
  for 3D simulations (filled points, red) and 1D ML2/$\alpha$ = 0.7
  model atmospheres (open points, black) at $\log g = 8$. The points
  are connected for clarity. \label{fg:f_vdyn}}
\end{figure}

The overshoot coefficients in Table~4 are limited by the $T_{\rm eff}$
range of our 3D simulations. Figure~\ref{fg:f_vdyn} compares the
maximum velocities, which peak slightly below the photosphere, for
$\langle {\rm 3D}\rangle$ and 1D ML2/$\alpha$ = 0.7 models at $\log g
= 8$. We applied the MLT parameterization that best represents the
maximum convective flux of the warmest 3D simulations (see
Figure~\ref{fg:f_fmax} and TB13c). The MLT suggests that velocities in
the photosphere for $14,000 < T_{\rm eff}$ (K) $ \lesssim 18,000$ are
still of the same order of magnitude as in cooler models, although the
upper $T_{\rm eff}$ limit depends critically on the MLT
parameterization. The large photospheric velocities are likely to
support strong overshoot layers in DA white dwarfs above our warmest
3D simulations, even though convection has a negligible effect on the
thermal structure.

 \begin{deluxetable}{llllllllll}
 \tabletypesize{\scriptsize}
 \tablecolumns{5}
 \tablewidth{0pt}
 \tablecaption{Overshoot Parameters for Closed 3D Simulations}
 \tablehead{
 \colhead{$T_{\rm eff}$} & 
 \colhead{$\log g$} & 
 \colhead{$v_{\rm base}$ {\tablenotemark{a}}} & 
 \colhead{$\log t_{\rm char}$ {\tablenotemark{b}}} & 
 \colhead{$\log H_{\rm P}$ {\tablenotemark{c}}} \\
 \colhead{(K)} & 
 \colhead{} & 
 \colhead{(10$^5$ cm s$^{-1}$)} & 
 \colhead{[s]} & 
 \colhead{[cm]} 
 }
 \startdata
  10540& 7.00&  3.36&  $-$0.28& 5.30\\
  11000& 7.00&  3.31&  $-$0.21& 5.34\\
  11501& 7.00&  3.52&  $-$0.29& 5.50\\
  12001& 7.00&  2.74&  $-$0.43& 5.54\\
  12501& 7.00&  1.56&  $-$0.49& 5.55\\
  13003& 7.00&  0.95&  $-$0.46& 5.51\\
  11498& 7.50&  3.27&  $-$0.50& 4.81\\
  11999& 7.50&  3.24&  $-$0.67& 4.91\\
  12500& 7.50&  3.10&  $-$0.86& 5.04\\
  13002& 7.50&  2.60&  $-$1.05& 5.10\\
  12099& 8.00&  3.33&  $-$0.96& 4.45\\
  12504& 8.00&  3.14&  $-$0.95& 4.35\\
  13000& 8.00&  3.04&  $-$1.15& 4.47\\
  13502& 8.00&  2.75&  $-$1.30& 4.56\\
  14000& 8.00&  1.88&  $-$1.37& 4.59\\
  12909& 8.50&  3.21&  $-$1.48& 3.98\\
  13453& 8.50&  2.92&  $-$1.44& 3.89\\
  14002& 8.50&  2.70&  $-$1.54& 3.99\\
  14492& 8.50&  2.49&  $-$1.68& 4.08\\
  14008& 9.00&  3.06&  $-$1.81& 3.48\\
  14591& 9.00&  2.71&  $-$1.85& 3.41\\
  14967& 9.00&  2.47&  $-$1.92& 3.47\\
 \enddata

\tablenotetext{a}{Corresponds to $\langle {\rm 3D}\rangle$ $v_{\rm z,rms}$ at the flux boundary.}
\tablenotetext{b}{Same as the decay time in Table A.1 of TL13c.}
\tablenotetext{c}{Corresponds to $\langle {\rm 3D}\rangle$ $P/(\rho g)$ at the flux boundary.}
\label{tb:t4}
\end{deluxetable}

\subsubsection{Open 3D Simulations}

For open 3D simulations, we can not directly extract quantities to
calibrate Eq.~(13). Furthermore, the assumption that the overshoot
timescale is the same as the surface granulation timescale is unlikely
to be valid, since the downdrafts have time for merging into the
hierarchical structure observed in simulations of deep, convective
envelopes \citep{nordlund09}. We propose instead that $t_{\rm char} =
H_{\rm v}/v_{\rm base}$, with the velocity scale height equal to the
pressure scale height as above. Therefore, $v_{\rm base}$ is the only
quantity that remains to be evaluated.

\begin{figure}[h!]
\epsscale{0.8}
\plotone{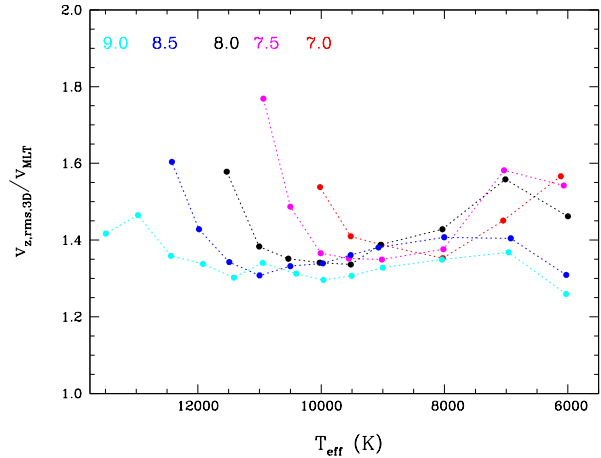}
\caption{Ratio of the 3D $v_{\rm z,rms}$ and 1D ML2/$\alpha_{\rm
    flux}$ velocities at $\log \tau_{\rm R }$ = 2.5 as a function of
  $T_{\rm eff}$ and $\log g$ (represented by different colors with the
  legend at the top). The points are connected for clarity. The
  ML2/$\alpha_{\rm flux}$ calibration is presented in Table
  2.\label{fg:f_vcalib}}
\end{figure}

For deep and essentially adiabatic convection zones, the MLT and 3D
simulations agree on the temperature gradient. An examination of the
MLT equations demonstrates that for very efficient convection ($F_{\rm
  conv} \sim F_{\rm tot}$), velocity is proportional to $\rho^{-1/3}$,
along with a dependence on heat capacity and molecular weight in the
presence of partial ionization. While the 1D velocity model is only an
idealization of the complex 3D dynamics, we suggest that the $v_{\rm
  3D}/v_{\rm 1D}$ ratio remains very similar across the deep
convection zone. This is seen in Figure~\ref{fg:f_vel} for the cooler
10,025~K model where convection is reasonably adiabatic below the
photosphere. Figure~\ref{fg:f_vcalib} shows the $\langle {\rm
  3D}\rangle$ versus 1D ML2$/\alpha_{\rm flux}$ velocity ratio for
open simulations and a reference layer identified by the criterion
$\log \tau_{\rm R} = 2.5$. This region is deep enough for convection
to be largely adiabatic, and far away from the bottom boundary to
prevent numerical effects. We observe small variations around a mean
value of $v_{\rm 3D}/v_{\rm 1D}$ = 1.5 for the DA white dwarfs with a
deep adiabatic convection zone. We suggest that this calibration
remains valid down to the bottom of the convection zone, as long as
$F_{\rm conv} \sim F_{\rm tot}$. We still face the problem, however,
that by definition $v_{\rm MLT,base} = 0$. We recommend instead to
take a characteristic velocity $v_{\rm MLT,base^{*}}$ one pressure
scale height above the bottom of the convection zone. In summary, for
the $T_{\rm eff}$ range below the one covered by Table~4, we propose
the following overshoot parameterization

\begin{eqnarray}
D_{\rm overshoot}(z) = ~~~~~~~~~~~~~~~~~~~~~~~~~~~~~~~ \nonumber \\
1.5 v_{\rm MLT,base^{*}} H_{\rm P} \exp(2(z-z_{\rm base})/H_{\rm P})~,
\end{eqnarray}

{\noindent}where all quantities are extracted from 1D
ML2$/\alpha_{\rm flux}$ structures as described above. 

We confirm the results of \citet{freytag96} that overshoot is
significant and present for all DA white dwarfs with convectively
unstable layers ($T_{\rm eff} \lesssim 18,000$~K). The total mass of
hydrogen included in the overshoot region may be a few orders of
magnitude greater than the mass included in the proper convection
zone. This effect is totally neglected in local MLT models, and our
proposed parameterization provides an order of magnitude estimate
(upper limit) of the overshoot velocities and macroscopic diffusion
coefficients. The resulting effects on the chemical abundances of
mixed elements, for instance in accreting white dwarfs in a steady
state, depend on the outcomes of chemical diffusion calculations.

\subsection{Improvements to the Local MLT}

The previous sections have revealed that the local MLT only depicts a
rough portrait of the underlying dynamical nature of convection, which
is illustrated by the need of having different parameterizations for
different applications. We note that non-local 1D MLT models could
provide a better match to the 3D results. In particular, the models
discussed in \citet{spiegel63}, \citet{skaley91}, \citet{dupret06},
and \citet{twocol} naturally deliver the Schwarzschild and flux
boundaries, as well as (partial) overshoot layers. In these non-local
MLT models, the more realistic physics is recovered at the expense of
adding more free parameters. In some sense, this is a more elegant and
accurate way of obtaining the Schwarzschild and flux boundaries than
we have proposed in this work. While it does require some
modifications of existing 1D model atmosphere and structure codes,
this should be investigated in the future.

\citet{montgomery04} have also presented a non-local convection model
for white dwarfs, although in this case it is not an extension of the
MLT theory. However, one issue for all non-local 1D models discussed
here is that they have not been very successful at modeling overshoot
velocities reproducing the exponential decay observed in RHD
simulations, which is the main part of the 1D models that we would
like to improve.

\subsection{ZZ Ceti Instability Strip}

The spectroscopically determined atmospheric parameters of pulsating
ZZ Ceti white dwarfs have been discussed in TL13c, as seen in the
light of our grid of $\langle {\rm 3D}\rangle$ spectra. We found that
the dominant 3D effect is on the spectroscopically determined surface
gravity, with an average shift of $\Delta \log g$ = $-$0.1 for ZZ Ceti
stars in the sample of \citet{gianninas11}. On the other hand, 3D
$T_{\rm eff}$ corrections depend critically on the calibration of the
MLT parameters in the reference 1D model atmospheres. Based on the 1D
ML2/$\alpha$ = 0.8 calibration, we observed a 3D shift of $\Delta
T_{\rm eff}$ = $-$225~K on average, although this is in the same range
as the uncertainties in the 3D corrections.  The {\it spectroscopic}
blue edge at $\log g = 8$, below which white dwarfs are pulsating, is
located at $T_{\rm eff} \sim 12,500$~K when relying on $\langle {\rm
  3D}\rangle$ spectra, while it is slightly warmer by 100~K based on
1D ML2/$\alpha$ = 0.8 model atmospheres. On the other hand, the
$\langle {\rm 3D}\rangle$ red edge is located at $T_{\rm eff} \sim
11,000$~K for $\log g = 8$. Overall, the observed position of the
instability strip is not changed significantly compared to earlier
investigations \citep{gianninas06,gianninas11}. We remind the reader
that the observed edges are defined from only a few pulsating and
constant objects, and that the individual errors on the spectroscopic
atmospheric parameters must also be considered.

Non-adiabatic asteroseismic models provide predictions for the
position of the blue edge of the ZZ Ceti instability strip, although
the results are highly sensitive to the parameterization of convection
\citep{fontaine94,gautschy96}. Recently, \citet{grootel12} relied on a
non-adiabatic code including time-dependent convection to study the
driving mechanism. Compared to earlier studies \citep[][and references
  therein]{fontaine08}, their approach neither assumes frozen
convection nor an instantaneous convection response during a pulsation
cycle. Using 1D ML2/$\alpha$ = 1.0 white dwarf structures similar to
those discussed in this work, they find a {\it seismic} blue edge at
$T_{\rm eff} = 11,970$~K for $\log g = 8$. Since the convective flux
contribution is critical in the non-adiabatic perturbation equations,
we can compare their results with our ML2/$\alpha_{\rm flux}$
calibration in Figure~\ref{fg:f_xmlt}.  We find that ML2/$\alpha_{\rm
  flux} \sim $ 1.0 at 12,000~K and $\log g = 8$, in very close
  agreement with the value generally used to predict the blue edge of
  the instability strip, based on seismic models. There seems to be a
slight discrepancy between the observed and predicted blue edges, the
latter being cooler by about $500$~K. We note, however, that the
current agreement is still fairly good considering the uncertainties
in the 3D simulations and spectroscopically determined atmospheric
parameters. It would be interesting to review the non-adiabatic
pulsation calculations with the new calibrated 1D envelopes or a
direct use of the $\langle {\rm 3D}\rangle$ convective flux profiles
\citep{gautschy96}. Finally, dynamical convection effects that are
missing from both current and newly calibrated 1D envelopes could also
have an impact on pulsations \citep{grootel12}.

At the red edge of the instability strip, \citet{grootel13} recently
revived an idea of \citet{hansen85} originally applied to the blue
edge.  They suggest that the red edge of the g-mode instabilities is
reached when the thermal timescale in the driving region (bottom of
the convection zone) becomes of the order of the pulsation period.
Beyond this limit, outgoing g-waves are no longer reflected back by
the atmospheric layers, and will lose their energy in the upper
atmosphere. Using this argument for g-modes of spherical-harmonic
degree $l$ = 1, the red edge lies at $\sim$11,000~K for $\log g = 8$
with ML2/$\alpha$ = 1.0 1D envelopes. In this range of $T_{\rm eff}$,
we predict a slightly shallower 3D convection zone, although it is
unlikely to impact in a qualitative way the results presented in
\citet{grootel13}.

\section{CONCLUSION}

We have presented a comparison of our grid of 3D RHD simulations for
70 DA white dwarfs, in the range $7.0 \leq \log g \leq 9.0$, with 1D
envelope models based on the mixing-length theory for
convection. While MLT only provides a bottom boundary of the
convection zone based on the Schwarzschild criterion, the 3D
stratifications are more complex. In 3D simulations, convective
structures are still accelerated just before reaching the
Schwarzschild boundary and the convective flux remains significant in
layers below the classical definition of the convection zone. In
addition, we confirm that DAs have strong lower overshoot layers,
where vertical velocities decay exponentially with a velocity scale
height of the order of the pressure scale height. 

We proposed two functions to calibrate ML2/$\alpha$ values in 1D
envelopes that best reproduce the 3D Schwarzschild and flux
boundaries, respectively, as a function of $T_{\rm eff}$ and $\log
g$. The calibration was performed from a direct comparison for
  closed simulations with shallow convection zones. For cool white
dwarfs with deep convection zones, the 3D simulations use an open
bottom boundary condition, and therefore do not include the lower part
of the convection zone. We rely on the fact that below the atmosphere,
upflows still evolve under adiabatic conditions. We have extracted the
3D asymptotic entropy values that correspond to the conditions in the
lower part of the convection zones, which were then employed to
calibrate ML2/$\alpha$ of 1D MLT envelopes.

We have found that for shallow and inefficient convection zones
($T_{\rm eff} \gtrsim 12,000$~K at $\log g = 8$), the MLT parameters
for the bottom of the convection zone poorly reproduce the overall
$\langle {\rm 3D}\rangle$ convective flux profile through the
convection zones. Mean 3D stratifications should be used for studies
that require detailed convective flux profiles.  For applications such
as chemical diffusion and convective mixing, the dominant convective
effect is likely to come from the overshoot velocities, which are
completely missing from local MLT envelopes. The extreme ratio between
convective and microscopic diffusion timescales prohibits the usage of
3D simulations to precisely calibrate the deep overshoot
layers. Instead, we reintroduce in the context of white dwarfs the
analytical overshoot parameterization initially proposed by
\citet{freytag96}, with new constraints based on the 3D
simulations. The next step will be to apply our calibrations to
non-adiabatic pulsation models as well as specific cases of white
dwarfs with convection zones contaminated by metals accreted from
former disrupted planets.

\acknowledgements

Support for this work was provided by NASA through Hubble Fellowship
grant \#HF-51329.01 awarded by the Space Telescope Science Institute,
which is operated by the Association of Universities for Research in
Astronomy, Inc., for NASA, under contract NAS 5-26555. This work was
supported by Sonderforschungsbereich SFB 881 "The Milky Way System"
(Subprojects A4 and A5) of the German Research Foundation (DFG).


\begin{thebibliography}{0}%
\makeatletter
\providecommand \@ifxundefined [1]{%
 \@ifx{#1\undefined}
}%
\providecommand \@ifnum [1]{%
 \ifnum #1\expandafter \@firstoftwo
 \else \expandafter \@secondoftwo
 \fi
}%
\providecommand \@ifx [1]{%
 \ifx #1\expandafter \@firstoftwo
 \else \expandafter \@secondoftwo
 \fi
}%
\providecommand \natexlab [1]{#1}%
\providecommand \enquote  [1]{``#1''}%
\providecommand \bibnamefont  [1]{#1}%
\providecommand \bibfnamefont [1]{#1}%
\providecommand \citenamefont [1]{#1}%
\providecommand \href@noop [0]{\@secondoftwo}%
\providecommand \href [0]{\begingroup \@sanitize@url \@href}%
\providecommand \@href[1]{\@@startlink{#1}\@@href}%
\providecommand \@@href[1]{\endgroup#1\@@endlink}%
\providecommand \@sanitize@url [0]{\catcode `\\12\catcode `\$12\catcode
  `\&12\catcode `\#12\catcode `\^12\catcode `\_12\catcode `\%12\relax}%
\providecommand \@@startlink[1]{}%
\providecommand \@@endlink[0]{}%
\providecommand \url  [0]{\begingroup\@sanitize@url \@url }%
\providecommand \@url [1]{\endgroup\@href {#1}{\urlprefix }}%
\providecommand \urlprefix  [0]{URL }%
\providecommand \Eprint [0]{\href }%
\providecommand \doibase [0]{http://dx.doi.org/}%
\providecommand \selectlanguage [0]{\@gobble}%
\providecommand \bibinfo  [0]{\@secondoftwo}%
\providecommand \bibfield  [0]{\@secondoftwo}%
\providecommand \translation [1]{[#1]}%
\providecommand \BibitemOpen [0]{}%
\providecommand \bibitemStop [0]{}%
\providecommand \bibitemNoStop [0]{.\EOS\space}%
\providecommand \EOS [0]{\spacefactor3000\relax}%
\providecommand \BibitemShut  [1]{\csname bibitem#1\endcsname}%
\let\auto@bib@innerbib\@empty
\end{thebibliography}%


\clearpage

\begin{thebibliography}{}

\bibitem[Asplund et al.(2009)]{asplund09} Asplund, M., Grevesse, N.,
  Sauval, A.~J., \& Scott, P.\ 2009, \araa, 47, 481

\bibitem[Bergeron et al.(1990)]{bergeron90} Bergeron, P., Wesemael,
  F., Fontaine, G., \& Liebert, J.\ 1990, \apjl, 351, L21

\bibitem[Bergeron et al.(1992)]{bergeron92} Bergeron, P., 
Wesemael, F., \& Fontaine, G.\ 1992, \apj, 387, 288 

\bibitem[Bergeron et al.(1995)]{bergeron95} Bergeron, P., Wesemael,
  F., Lamontagne, R., et al.\ 1995, \apj, 449, 258

\bibitem[B{\"o}hm(1963)]{bohm63} B{\"o}hm, K.-H.\ 1963, \apj, 
138, 297 

\bibitem[B{\"o}hm(1968)]{bohm68} B{\"o}hm, K.-H.\ 1968, \apss, 2, 375 

\bibitem[Bohm \& Cassinelli(1971)]{bohm71} B{\"o}hm, K.~H., \& Cassinelli,
  J.\ 1971, \aap, 12, 21

\bibitem[B\"ohm-Vitense(1958)]{MLT} B\"ohm-Vitense, E.\ 1958, ZAp, 46,
  108

\bibitem[Brassard \& Fontaine(1997)]{brassard97} Brassard, P.,~\&
  Fontaine. G. 1997, in White dwarfs, ed. J. Isern, M. Hernanz, \&
  E. Garcia-Berro (Dordrecht: Kluwer Academic Publishers), 214, 451

\bibitem[Caffau et al.(2011)]{caffau11} Caffau, E., Ludwig, H.-G.,
  Steffen, M., Freytag, B., \& Bonifacio, P.\ 2011, \solphys, 268, 255

\bibitem[Chan \& Sofia(1989)]{chan89} Chan, K.~L., \& Sofia, S.\ 1989,
  \apj, 336, 1022

\bibitem[Chan \& Gigas(1992)]{chan92} Chan, K.~L., \& Gigas, D.\ 1992,
  \apjl, 389, L87

\bibitem[Chen \& Hansen(2011)]{chen11} Chen, E.~Y., \& Hansen,
  B.~M.~S.\ 2011, \mnras, 413, 2827

\bibitem[Dupret et al.(2006)]{dupret06} Dupret, M.-A., Goupil, M.-J.,
  Samadi, R., Grigahc{\`e}ne, A., \& Gabriel, M.\ 2006, in Proceedings
  of SOHO 18/GONG 2006/HELAS I, Beyond the spherical Sun,
  ed. M. Thompson (Noordwijk, ESA), 624, 78

\bibitem[Dupuis et al.(1993)]{dupuis93} Dupuis, J., Fontaine, 
G., Pelletier, C., \& Wesemael, F.\ 1993, \apjs, 84, 73 

\bibitem[Fontaine \& Brassard(2008)]{fontaine08} Fontaine, G., \&
  Brassard, P.\ 2008, \pasp, 120, 1043
  
\bibitem[Fontaine \& van Horn(1976)]{fontaine76} Fontaine, G., \& van
  Horn, H.~M.\ 1976, \apjs, 31, 467

\bibitem[Fontaine et al.(1994)]{fontaine94} Fontaine, G., 
Brassard, P., Wesemael, F., \& Tassoul, M.\ 1994, \apjl, 428, L61 

\bibitem[Fontaine et al.(2001)]{fontaine01} Fontaine, G., Brassard,
  P., \& Bergeron, P. 2001, \pasp, 113, 409
  
  \bibitem[Freytag 
\& Salaris(1999)]{freytag99} Freytag, B., \& Salaris, M.\ 1999, \apjl, 513, L49 

\bibitem[Freytag et al.(1996)]{freytag96} Freytag, B., Ludwig, H.-G.,
  \& Steffen, M.\ 1996, \aap, 313, 497
  
\bibitem[Freytag et al.(2010)]{freytag10} Freytag, B., Allard, F.,
  Ludwig, H.-G., Homeier, D., \& Steffen, M.\ 2010, \aap, 513, A19
  
\bibitem[Freytag et al.(2012)]{freytag12} Freytag, B., Steffen, M.,
  Ludwig, H.-G., et al.\ 2012, Journal of Computational Physics, 231,
  919

\bibitem[Gianninas et al.(2006)]{gianninas06} Gianninas, A., Bergeron,
  P., \& Fontaine, G.\ 2006, \aj, 132, 831

\bibitem[Gianninas et al.(2011)]{gianninas11} Gianninas, A., Bergeron,
  P., \& Ruiz, M.~T.\ 2011, \apj, 743, 138

\bibitem[Gautschy et al.(1996)]{gautschy96} Gautschy, A., Ludwig,
  H.-G., \& Freytag, B.\ 1996, \aap, 311, 493

\bibitem[van Grootel et al.(2012)]{grootel12} van Grootel, V., Dupret,
  M.-A., Fontaine, G., et al.\ 2012, \aap, 539, A87

\bibitem[van Grootel et al.(2013)]{grootel13} van Grootel, V.,
  Fontaine, G., Brassard, P., \& Dupret, M.-A.\ 2013, \apj, 762, 57

\bibitem[Hansen(1999)]{hansen99} Hansen, B.~M.~S.\ 1999, \apj, 520,
  680
  
\bibitem[Hansen et al.(1985)]{hansen85} Hansen, C.~J., Winget, 
D.~E., \& Kawaler, S.~D.\ 1985, \apj, 297, 544 

\bibitem[Hummer \& Mihalas(1988)]{hm88} Hummer, D.~G., \& Mihalas,
  D.\ 1988, \apj, 331, 794

\bibitem[Koester(2009)]{koester09} Koester, D.\ 2009, \aap, 498, 517

\bibitem[Koester et al.(1994)]{koester94} Koester, D., Allard, N.~F.,
  \& Vauclair, G.\ 1994, \aap, 291, L9

\bibitem[Ludwig et al.(1999)]{ludwig99} Ludwig, H.-G., Freytag, B., \&
  Steffen, M.\ 1999, \aap, 346, 111
  
\bibitem[Ludwig et al.(2002)]{ludwig02} Ludwig, H.-G., Allard, F., \&
  Hauschildt, P.~H.\ 2002, \aap, 395, 99

\bibitem[Ludwig et al.(2008)]{ludwig08} Ludwig, H.-G., Caffau, E.,
    \& Ku{\v c}inskas, A.\ 2008, In IAU Symp. 252, ed. L. Deng \&
    K. L. Chan (Cambridge: Cambridge Univ. Press), 75

\bibitem[Montgomery \& Kupka(2004)]{montgomery04} Montgomery, M.~H.,
  \& Kupka, F.\ 2004, \mnras, 350, 267

\bibitem[Nordlund et al.(2009)]{nordlund09} Nordlund, {\AA}., 
Stein, R.~F., \& Asplund, M.\ 2009, Living Reviews in Solar Physics, 6, 2 

\bibitem[Paquette et al.(1986)]{paquette86} Paquette, C., Pelletier,
  C., Fontaine, G., \& Michaud, G.\ 1986, \apjs, 61, 197

\bibitem[Pelletier et al.(1986)]{pelletier86} Pelletier, C., 
Fontaine, G., Wesemael, F., Michaud, G., 
\& Wegner, G.\ 1986, \apj, 307, 242 

\bibitem[Renedo et al.(2010)]{renado10} Renedo, I., Althaus, L.~G.,
  Miller Bertolami, M.~M., et al.\ 2010, \apj, 717, 183

\bibitem[Salaris et al.(2010)]{salaris10} Salaris, M., Cassisi, S.,
  Pietrinferni, A., Kowalski, P.~M., \& Isern, J.\ 2010, \apj, 716,
  1241

\bibitem[Saumon et al.(1995)]{saumon95} Saumon, D., Chabrier, 
G., \& van Horn, H.~M.\ 1995, \apjs, 99, 713 

\bibitem[Scott et al.(2014a)]{scott14a} Scott, P., Grevesse,
  N., Asplund, M., et al.\ 2014a, arXiv:1405.0279

\bibitem[Scott et al.(2014b)]{scott14b} Scott, P., Asplund,
  M., Grevesse, N., Bergemann, M., \& Sauval, A.~J.\ 2014b,
  arXiv:1405.0287

\bibitem[Skaley 
\& Stix(1991)]{skaley91} Skaley, D., \& Stix, M.\ 1991, \aap, 241, 227 

\bibitem[Spiegel(1963)]{spiegel63} Spiegel, E.~A.\ 1963, \apj, 
138, 216 

\bibitem[Steffen(1993)]{steffen93} Steffen, M.\ 1993, in Inside the
  stars, ed. W. Weiss \& A. Baglin, (San Francisco, ASP), 40, 300

\bibitem[Stein \& Nordlund(1989)]{stein89} Stein, R.~F., \& Nordlund,
  A.\ 1989, \apjl, 342, L95

\bibitem[St{\"o}kl(2008)]{twocol} St{\"o}kl, A.\ 2008, \aap, 490, 1181 

\bibitem[Tassoul et al.(1990)]{tassoul90} Tassoul, M., Fontaine, G.,
  \& Winget, D.~E.\ 1990, \apjs, 72, 335

\bibitem[Tremblay et al.(2011)]{tremblay11} Tremblay, P.-E.,
  Bergeron, P. \& Gianninas, A.\ 2011, \apj, 730, 128

\bibitem[Tremblay et al.(2013a)]{tremblay13a} Tremblay, P.-E., Ludwig,
  H.-G., Steffen, M., \& Freytag, B.\ 2013a, \aap, 552, A13 (TL13a)

\bibitem[Tremblay et al.(2013b)]{tremblay13b} Tremblay, P.-E., Ludwig,
  H.-G., Freytag, B., Steffen, M., \& Caffau, E.\ 2013b, \aap, 557, A7
  (TL13b)

\bibitem[Tremblay et al.(2013c)]{tremblay13c} Tremblay, P.-E., Ludwig,
  H.-G., Steffen, M., \& Freytag, B.\ 2013c, \aap, 559, A104 (TL13c)

\bibitem[Unno(1957)]{unno57} Unno, W.\ 1957, \apj, 126, 259 

\bibitem[Wedemeyer-B{\"o}hm \& Rouppe van der
  Voort(2009)]{wedemeyer09} Wedemeyer-B{\"o}hm, S., \& Rouppe van der
  Voort, L.\ 2009, \aap, 503, 225

\bibitem[Zahn(1991)]{zahn91} Zahn, J.-P.\ 1991, \aap, 252, 179 

\end{thebibliography}
\end{document}